\documentclass[twocolumn]{IEEEtran}

\IEEEoverridecommandlockouts
\usepackage{cite}
\usepackage{amsmath,amssymb,amsfonts}
\usepackage{algorithmic}
\usepackage{graphicx}
\usepackage{textcomp}
\usepackage{xcolor}
\usepackage{bm}
\usepackage{stfloats}
\usepackage{mathrsfs}
\usepackage{algorithm}
\def\BibTeX{{\rm B\kern-.05em{\sc i\kern-.025em b}\kern-.08em
    T\kern-.1667em\lower.7ex\hbox{E}\kern-.125emX}}
\begin{document}

\title{Precoding  and Transmit Antenna Subarray Selection for Secure Hybrid Spatial Modulation\\
}

\author{Feng Shu,~Xiaoyu Liu,~Xinyi~Jiang,~Ling~Xu,~Guiyang Xia,~and~Jiangzhou Wang}

\maketitle

\begin{abstract}
Spatial modulation (SM) is a particularly important form of multiple-input-multiple-output (MIMO). Unlike traditional MIMO, it uses both modulation symbols and antenna indices to carry information. In this paper, to avoid the high cost and circuit complexity of fully-digital SM, we mainly consider the hybrid SM system with a hybrid precoding transmitter architecture, combining a digital precoder and an analog precoder. Here, the partially-connected structure is adopted with each radio frequency chain (RF) being connected to a transmit antenna subarray (TAS). In such a system, we made an investigation of  secure hybrid precoding and transmit antenna subarray selection (TASS) methods. Two hybrid precoding methods, called maximizing the approximate secrecy rate (SR) via gradient ascent (Max-ASR-GA) and  maximizing the approximate SR via alternating direction method of multipliers (Max-ASR-ADMM), are  proposed  to improve the SR performance.  As for TASS, a high-performance method of maximizing the approximate SR (Max-ASR) TASS method is first presented. To reduce its high complexity, two low-complexity TASS methods, namely maximizing the eigenvalue (Max-EV) and  maximizing the product of signal-to-interference-plus-noise ratio and artificial noise-to-signal-plus-noise ratio (Max-P-SINR-ANSNR), are proposed.  Simulation results will demonstrate that  the proposed Max-ASR-GA and Max-ASR-ADMM hybrid precoders harvest substantial SR performance gains over existing method. For TASS, the proposed three methods Max-ASR, Max-EV, and Max-P-SINR-ANSNR perform better than existing leakage method. Particularly, the proposed Max-EV and Max-P-SINR-ANSNR  is low-complexity at the expense of a little performance loss compared with Max-ASR.
\end{abstract}
%In contrast to conventional SM systems, the performance of hybrid SM systems can be enhanced by selecting a TAS rather than a single transmit antenna based on spatial bits to send modulation bits.

\section{Introduction}
As an essential key technique for  future generation wireless communications like 6G and WiFi 7, multiple-input-multiple-output (MIMO)  has obtained a continuous extensive and intensive huge amount of research activities  from the end of the last century, which can greatly improve the system performance in wireless communications. Two typical forms of MIMO are bell laboratories layer space-time (BLAST) in\cite{Foschini2010Layered} and space time coding (STC) in \cite{Yu2015Power}. They make their effort to strike a good balance between spatial multiplexing and diversities in MIMO systems. However,
as the third form of MIMO, spatial modulation (SM) is attracting more and more research attention from industry and academia due to its advantages of no inter-channel interference (ICI) and inter-antenna synchronization (IAS) \cite{Renzo2011Spatial}. It works by mapping a block of information bits into two information-carrying units. The first information carrying unit is a amplitude/phase modulation (APM) symbol chosen from the signal-constellation diagram. The second information-carrying unit is a transmit antenna index chosen, while other transmit antennas are not activated\cite{Mesleh2008Spatial,Renzo2011Spatial,Jeganathan2012Spatial}.

There are several ways to improve the performance of SM including transmit antenna selection\cite{Rajashekar2013Antenna,Xia2018AS,Xia2019AN}, linear precoding \cite{Jin2015Linear,Yang2011Transmitter}, power allocation\cite{shu2019high,xia2018power} and so on. The physical layer security of conventional SM  is an urgent problem. The authors in \cite{Aghdam2015On} focused their attention on the secrecy behavior of SSK and SM, and derived the expression for secrecy rate (SR). In \cite{Wang2015Secrecy}, the secrecy performance of SM  was improved by making a combination of useful signal and jamming against unknown eavesdropping, and the corresponding SR performance is analyzed. The authors in \cite{Wang2016Spatial} redefined the transmit antenna indices from the viewpoint of channel state information (CSI) to prevent eavesdropping by assuming that  the imperfect legal CSI is available for the eavesdropper.  Then, a full-duplex receiver assisted secure spatial modulation  scheme was proposed in \cite{Liu2017Secure}. Here, jamming signals are designed to protect legitimate receivers from self-interference (SI) and to interfere eavesdroppers. Besides, in \cite{Shu2018two}, the authors proposed two secure transmit antenna selection procedures, which can achieve a better secrecy performance. The  problem of power allocation (PA) between confidential message and artificial noise (AN)\cite{zhao2017artificial,zheng2015multi} was addressed in \cite{shu2019high}. And the authors in\cite{shu2019high,xia2018power} have proposed two PA strategies with  performance approaching that of the optimal PA factor.

It is very important to study the security of multiple-antenna communication systems\cite{chen2016survey,hu2016robust,shu2016robust,Zhou2019UAV,zou2016physical}. Furthermore, the security of SM systems is also very important. The precoding  schemes of traditional SM can also be generalized to secure SM. For most  conventional PSM schemes, the antenna indices at  receiver is utilized to carry bits rather than the antenna indices of the transmitter as spatial bits. The optimization of the precoding matrix at transmitter is to address the issues of preprocessing and detection of PSM signals at receiver in order to improve bit-error-rate (BER) performance \cite{Yang2011Transmitter}. However, the authors in \cite{Wu2015Secret} have proposed a time-varying precoder for the secret PSM (SPSM), which generated a time-varying interference to the eavesdropper and retains all  advantages of PSM at the legal receiver. Then they  also derived  the upper bounds for BERs at legal receiver and  eavesdropper in the massive MIMO systems in \cite{Wu2015Secure}, and designed a precoder by jointly minimizing the receive power at eavesdropper and maximizing the receive power at legal receiver in \cite{Wu2016Transmitter}. The kind of PSM is also extended to secure multiuser MIMO downlink scenario by introducing a scrambling matrix to disturb the eavesdropper\cite{Chen2016Secure}.

Until now, we make a literature review about fully-digital (FD) SM. However, as the number of transmit antennas tends to medium-scale or large-scale, the circuit cost and complexity will become a burden on SM. To deal with this problem, hybrid SM, combining hybrid MIMO structure in \cite{Gao2016Energy,shu2018low} and SM, emerges as the times require\cite{Cui2016Hybrid,Y2017Hybrid,Sheng2018Macro,lee2017adaptive,He2017Spectral,lu2018low,he2018spatial}. It could make a good balance among cost, complexity and performance.

 %In order to improve the performance of SM systems, in addition to PSM, there are hybrid precoders in hybrid SM systems. In hybrid structures, the data stream is first processed in a digital precoder, and then flows through a few number of RF chains into an analog precoder, which is an analog phase-shifting network composed of phase shifters (PSs). The hybrid transmitter consists of two different structures: fully-connected architecture, in which each RF chain is connected to all transmit antennas via PSs, and partially-connected architecture\cite{shu2018low}, in which each RF chain is connected to only a transmit antenna subarray (TAS).

In the case of hybrid partially-connected architecture, the total antenna array is divided into multiple transmit antenna subarrays (TASs) with each being connected to single RF chain. Thus, the spatial bits can be carried by selecting the indices of TASs rather than those of transmit antennas. However, there were several papers focusing on hybrid SM. The hybrid SM was first proposed in \cite{Cui2016Hybrid}. But in \cite{Cui2016Hybrid}, it actually used analog precoding, and its digital precoder was replaced by the process of TAS selection (TASS) in SM. Besides, hybrid  SM was also extended to the multi-user scenario in \cite{Y2017Hybrid}, which showed that the hybrid SM system with hybrid beamforming at transmitter and digital combining at the receiver can achieve an excellent BER performance.

The authors in \cite{Sheng2018Macro} made an investigation of the macro SM, in which the indexes of the small base stations were used to carry information bits and they also proposed a low-complexity detection method. A multimode hybrid precoder is designed for proposed analog precoding-aided virtual SM and obtain BER performance gain compared with the conventional precoding-aided MIMO systems\cite{lee2017adaptive}.

For hybrid generalized SM (GSM), in \cite{He2017Spectral}, an analog precoder of maximizing spectrum efficiency (SE) was designed with the verified superiority in SE performance. In \cite{lu2018low}, the digital and analog precoders by turbo optimization was proposed for hybrid GSM systems to improve its SE performance. Furthermore, in \cite{he2018spatial}, the closed-form expression of the achievable  SE lower bound of was derived firstly. Then, via exploiting the concavity of the  SE expression in digital precoding vectors, the digital precoding vectors were computed, the convex relaxation was invoked to handle the non-convex constraints of analog precoding, and the corresponding problem of  analog precoding was converted into a convex optimization problem. Finally, the designed digital and analog precoding schemes  harvested  higher  SE gains over existing methods.

 In this paper, several techniques  in this paper will be developed  to improve the secrecy performance of hybrid SM as follows:  TASS, digital precoding for confidential message (CM), AN projection, and analog precoding.  Our main contributions  are summarized as follows:

%Only \cite{Tian2017Hybrid} has considered the design of secure hybrid precoding in the case of known eavesdropper's channel and unknown eavesdropper's channel. The authors in \cite{ramadan2017hybrid} has designed hybrid precoders for physical-layer security in multiple input single output (MISO) orthogonal frequency division multiplexing (OFDM) systems. In the system model of this paper, the transmitted signal is also assisted by AN, and we have designed a hybrid precoding based on SR and compares it with conventional hybrid precoding. Besides, in this paper, the secure TASS schemes for this system based on SR are also proposed.

\begin{enumerate}
 \item  A hybrid secure SM system model is established, where the transmitter is equipped with the partially-connected architecture, and the desired receiver and eavesdropping receiver are equipped with fully digital architecture. Since each RF chain is connected to a TAS in the partially-connected architecture, the spatial bits are carried by activating the TAS rather than a single transmit antenna, while the APM symbols are transmitted by the active TAS. Each TAS carries single bit stream by using multiple antennas. This will create spatial diversity and will be exploited to improve the secure performance of SM. In addition, with the help of AN  interfering with the eavesdropper, the useful signal is sent along the desired subspace  to further improve the security of this system.
 \item  To improve the secrecy performance, two hybrid precoding methods, maximizing the approximate SR (ASR) based on the gradient ascent (Max-ASR-GA) and maximizing the approximate SR (Max-ASR) based on alternating direction method of multipliers (Max-ASR-ADMM), are proposed. Due to the fact that there is no closed-form expression of SR, an ASR expression is presented for hybrid precoding design. The proposed Max-ASR-ADMM method can be expressed as a general form consensus problem, which can be addressed by exiting ADMM. Simulation results show that the SR performance of the proposed Max-ASR-GA is better than  conventional hybrid precoding method: SDR-AltMin and the proposed Max-ASR-ADMM. In particular, the proposed  Max-ASR-ADMM has a better SR performance than SDR-AltMin in the medium and high regions.
    % For Max-ASR-GA method,  the gradient of the approximate SR to the total precoding matrix (i.e., the product of digital precoding matrix and analog precoding matrix) is calculated. Then the Max-ASR-GA method can be used to obtain the secure precoding, and then the digital precoding and analog precoding are obtained.
 \item  Generally, the number of TASs is not a power of two, so TASS is required. In order to improve the secrecy performance, three secure TASS schemes are proposed. Maximizing approximate SR (Max-ASR) TASS method is proposed to consider the effect of the above two factors on secrecy performance.  To reduce the complexity of Max-ASR, the maximizing the eigenvalue (Max-EV) TASS method and the maximizing the product of signal-to-interference-plus-noise ratio and artificial noise-to-signal-plus-noise ratio (Max-P-SINR-ANSNR) are proposed as two low complexity methods. Since each TAS corresponds to a sub-channel, it can achieve a good rate performance by  choosing the TASs corresponding to the sub-channels with large channel gains. The eigenvalues of each sub-channel are sorted in order and the corresponding TASs of the sub-channels with large eigenvalues are used.  The proposed Max-EV performs slightly better than the proposed Max-P-SINR-ANSNR in the medium and high SNR regions with the same complexity as the latter in terms of SR. The proposed Max-ASR harvests a substantial SR performance gain over Max-EV and Max-P-SINR-ANSNR.

    \end{enumerate}

\section{System Model}

\begin{figure*}[t]
\centerline{\includegraphics[width=1\textwidth]{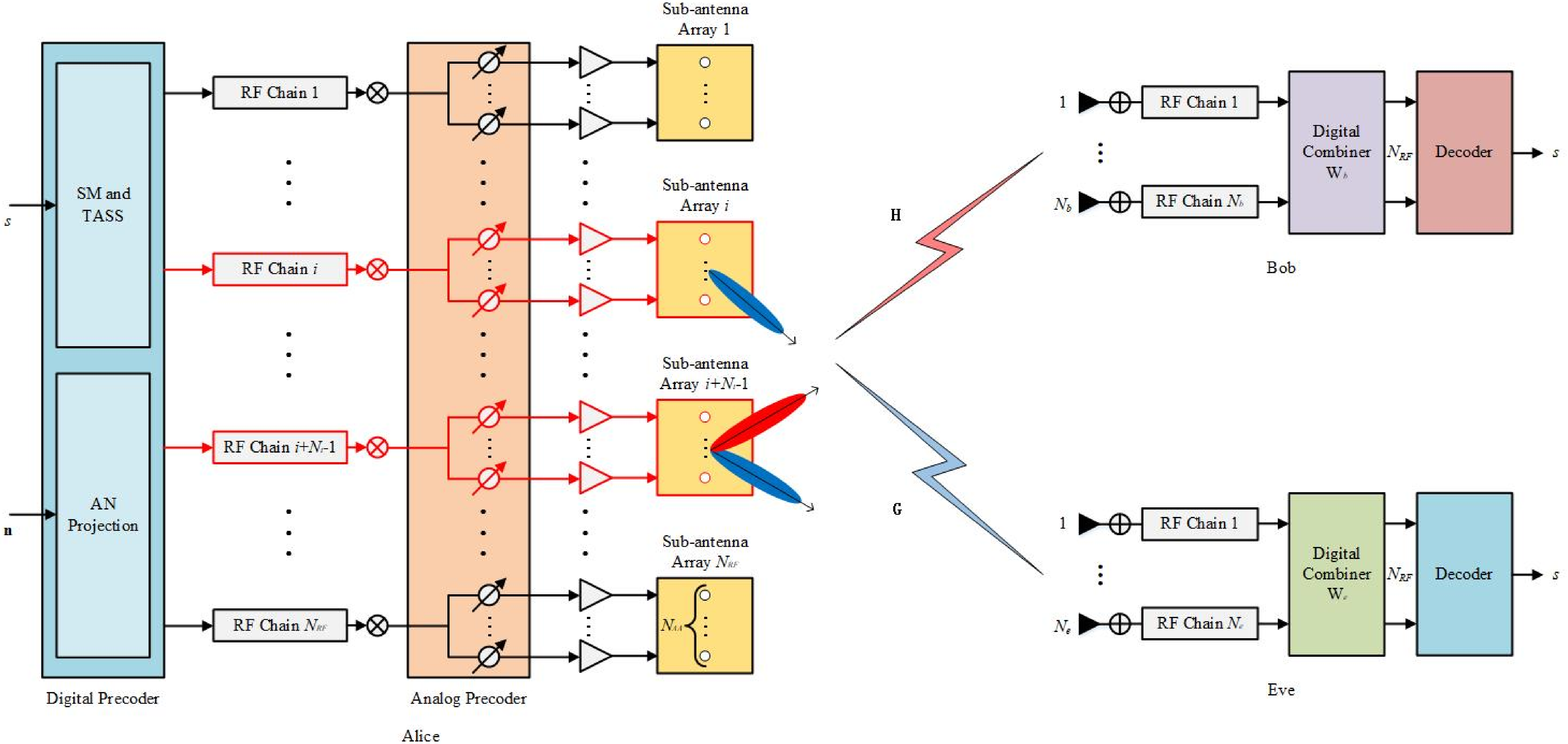}}
\caption{Block diagram for secure hybrid precoding SM system.}
\label{fig1}
\end{figure*}

Consider a system of hybrid SM as shown in Fig.~\ref{fig1}, where the transmitter (Alice) is equipped with $N_{\rm{RF}}$ RF chains with each having $N_{\rm{AA}}$ antennas to transmit one data stream to the legitimate receiver (Bob) equipped with $N_{b}$ antennas and $N_b^{\rm{RF}}$ RF chains. There is an eavesdropper (Eve) with $N_{e}$ antennas and $N_e^{\rm{RF}}$ RF chains trying to eavesdrop on the data stream from Alice to Bob. Alice uses the hybrid sub-connected structure, while Bob and Eve use fully digital structure, namely $N_{b}\!=\!N_b^{\rm{RF}}$ and $N_{e}\!=\!N_e^{\rm{RF}}$.
%And we assume Alice, Bob and Eve adopt uniform linear arrays (ULA).

 Now, we extend conventional SM to hybrid structure transmitters in MIMO systems. Therefore, in Fig.~\ref{fig1} system, the first part information bits are used to choose one of TASs and the second part information bits are mapped to the modulation symbol.
%In general, we assume that $N_a$ AA is selected to transmit useful information, so there are $\kappa=\begin{pmatrix} N_t \\ N_a \end{pmatrix}$ number of possible choices. Therefore, $\lfloor\log_2{\kappa}\rfloor$ spatial bits can be transmitted.
In what follows, it is assumed that $N_t=\lfloor N_{\rm{RF}}\rfloor_2$ due to the reason that the number of TASs is confined to a power of two. Let $I_{\rm{ASS}}=\{l_i\}_{i=1}^{N_t}$ be the legitimate TAS subset, where the element $l_i$ is selected from set $\left\{1,2,\cdots,N_{\rm{RF}}\right\}$. A proper TASS strategy  can clearly improve system performance.
$\mathbb{I}$ is the set of enumerations of all possible  $Q=\begin{pmatrix} N_{\rm{RF}} \\ N_t \end{pmatrix}$ combinations of TAS subsets $I_{\rm{ASS}}$.
The modulation symbol first experiences the digital precoder and then passes through the RF chain connected by the active TAS.
%The SM and TAAS processes described above are implemented in the digital precoder.
Considering the security of communication,  AN is needed to disturb the eavesdropper as an efficient secure tool. AN is also projected in digital precoding stage. After the precoded CM plus the projected AN passing the corresponding RF chains,and the phase shifters (PSs) of analog precoding and AN. Finally, the transmit signal is in the following form
\begin{align} \label{x} \nonumber
\textbf{s}&=\sqrt{\beta P}\textbf{E}_i\textbf{F}_{\rm{RF}}\textbf{F}_{\rm{BB}}b_j+\sqrt{(1-\beta)P}\textbf{TF}_{\rm{RF}}\textbf{T}_{\rm{BB}}\textbf{n}\\
&=\sqrt{\beta P}\textbf{E}_i\textbf{P}b_j+\sqrt{(1-\beta)P}\textbf{TP}_{\rm{AN}}\textbf{n},
\end{align}
where $b_j$ is the digital symbol chosen from the $M$-ary constellation for $j\in\mathbb{M}=\left\{1,2,\cdots,M\right\}$, and satisfies $\mathbb{E}|b_j|^2=1$. $\textbf{E}_i\!=\!\textrm{diag}[\textbf{0},\cdots\!,\textbf{0},\textbf{e}_i,\textbf{0},\cdots\!,\textbf{0}]\!\in\! \mathbb{R}^{N_{\rm{AA}}N_{\rm{RF}} \times N_{\rm{AA}}N_{\rm{RF}}}$ is a block diagonal matrix, and  its $i$-th sub-block is the identity matrix, that is, $\textbf{e}_{i}=\textbf{I}_{N_{\rm{AA}}}$, which means that the $i$-th TAS is activated with $i\in I_{\rm{ASS}}$.
$\textbf{T}=\textrm{diag}[\textbf{t}_1,\textbf{t}_2,\cdots,\textbf{t}_{N_{\rm{RF}}}]$ is a block diagonal TASS matrix similar to $\textbf{E}_i$, where $\textbf{t}_i=\textbf{I}_{N_{\rm{AA}}}$ for all $i\in I_{\rm{ASS}}$, and the other sub-blocks are all-zero matrix, which means that $N_t$ TASs are selected. Since there are Q combinations for $I_{\rm{ASS}}\in\mathbb{I}$, i.e., there are Q possibilities of $\textbf{T}$, which can be expressed as $\textbf{T}_1,\textbf{T}_2,\ldots,\textbf{T}_Q$.
$\textbf{n}\sim\mathcal{CN}(0,\textbf{I}_{N_t})$ is the  AN vector, and $\beta$ is the power allocation factor between CM and AN. $P$ is the transmit power. And the digital precoding matrix of CM is given by
\begin{align} \label{F_bb}
\textbf{F}_{\rm{BB}}=[p_1 \ p_2 \ \cdots \ p_{N_{\rm{RF}}}]^T,
\end{align}
and $\textbf{T}_{\rm{BB}}\!\in\! \mathbb{C}^{N_{\rm{RF}} \times N_t}$ is the digital projection matrix of AN, also called the AN precoding matrix. Since only $N_t$ TASs are selected from $N_{\rm{RF}}$ TASs to work, and the AN $\textbf{n}$ is only projected to the selected $N_t$ TASs, so $\textbf{T}_{\rm{BB}}$ only has the corresponding $N_t$ rows with non-zero vectors and the remaining rows are all $\textbf{0}$. After passing through the corresponding RF chain, the precoded CM plus AN from each RF chain is delivered to the corresponding $N_{\rm{AA}}$ PSs to perform the corresponding analog precoding. This process can be denoted by the $N_{\rm{AA}}N_{\rm{RF}} \times N_{\rm{RF}}$ analog precoding matrix $\textbf{F}_{\rm{RF}}$ as follows
\begin{equation} \label{F_rf}
\textbf{F}_{\rm{RF}}={
\left[
\begin{array}{cccc}
    \textbf{f}_1 & \textbf{0} & \cdots & \textbf{0} \\
    \textbf{0} & \textbf{f}_2 & \cdots & \textbf{0} \\
    \vdots & \vdots & \ddots & \vdots \\
    \textbf{0} & \textbf{0} & \cdots & \textbf{f}_{N_{\rm{RF}}}
\end{array}
\right ]_{N_{\rm{AA}}N_{\rm{RF}} \times N_{\rm{RF}}}},
\end{equation}
where $\textbf{f}_i\in \mathbb{C}^{N_{\rm{AA}} \times 1}$ is the $i$th analog weighting vector for $i=1,2,\cdots,N_{\rm{RF}}$, and all its elements have the same amplitude $1/\sqrt{N_{\rm{AA}}}$ but distinct phases. Therefore, $\textbf{P}=\textbf{F}_{\rm{RF}}\textbf{F}_{\rm{BB}}$ can be expressed as
\begin{align}\label{P-def}
\textbf{P}=\left[\textbf{p}_1^T \ \textbf{p}_2^T \ \ldots \ \textbf{p}_{N_{\rm{RF}}}^T\right]^T,
 \end{align}
 where $\textbf{p}_i\!\in\! \mathbb{C}^{N_{\rm{AA}} \times 1},i\!=\!1,2,\cdots,N_{\rm{RF}}$, and $\textbf{P}_{\rm{AN}}\!=\!\textbf{F}_{\rm{RF}}\textbf{T}_{\rm{BB}}$.
The associated transmit vector after precoding and TASS can be expressed as
\begin{align}\label{x-def}
\textbf{x}=\textbf{E}_i\underbrace{\textbf{F}_{\rm{RF}}\textbf{F}_{\rm{BB}}}_{\textbf{P}}b_j=
[\underbrace{\textbf{0}\ \cdots\ \textbf{0}}\limits_{i-1}\ \textbf{b}_j\ \underbrace{\textbf{0}\ \cdots\ \textbf{0}}\limits_{N_{\rm{RF}}-i}]^T,
\end{align}
where $\textbf{b}_j\in \mathbb{C}^{1\times N_{\rm{AA}}}$ is the modulation symbol $b_j$ after hybrid precoding, and $i\in I_{\rm{ASS}}$ denotes the index of the activated TAS.
The received signals at Bob and Eve can be represented as
\begin{align} \label{rb} \nonumber
\textbf{r}_b&=\sqrt{\beta P}\textbf{HE}_i\textbf{F}_{\rm{RF}}\textbf{F}_{\rm{BB}}b_j \\
&+ \sqrt{(1-\beta)P}\textbf{HT}\textbf{F}_{\rm{RF}}\textbf{T}_{\rm{BB}}\textbf{n}+\textbf{n}_b,
\end{align}
%=&\sqrt{\beta P}\textbf{h}_i\textbf{f}_ip_ib_j+
%\sqrt{(1-\beta)P}\textbf{H}_\textbf{T}\textbf{A}_{\textbf{T}}\textbf{D}_{n,\textbf{T}}\textbf{n}+\textbf{n}_b
\begin{align} \label{re} \nonumber
\textbf{r}_e&=\sqrt{\beta P}\textbf{GE}_i\textbf{F}_{\rm{RF}}\textbf{F}_{\rm{BB}}b_j \\
&+ \sqrt{(1-\beta)P}\textbf{GT}\textbf{F}_{\rm{RF}}\textbf{T}_{\rm{BB}}\textbf{n}+\textbf{n}_e,
\end{align}
%=&\sqrt{\beta P}\textbf{g}_i\textbf{f}_ip_ib_j+
%\sqrt{(1-\beta)P}\textbf{G}_\textbf{T}\textbf{A}_{\textbf{T}}\textbf{D}_{n,\textbf{T}}\textbf{n}+\textbf{n}_b
where $\textbf{H}\in \mathbb{C}^{N_b \times N_{\rm{AA}}N_{\rm{RF}}}$ is the channel matrix of Alice-to-Bob which can be thought as $\textbf{H}=[\textbf{h}_1,\textbf{h}_2,\cdots,\textbf{h}_{N_{\rm{RF}}}]$ and $\textbf{h}_i\in \mathbb{C}^{N_{b} \times N_{\rm{AA}}}$ is the channel matrix from TAS $i$ to Bob.
$\textbf{G}\in \mathbb{C}^{N_e \times N_{\rm{AA}}N_{\rm{RF}}}$ is the channel matrix from Alice to Eve and $\textbf{G}$ can be given by $\textbf{G}=[\textbf{g}_1,\textbf{g}_2,\cdots,\textbf{g}_{N_{\rm{RF}}}]$ and $\textbf{g}_i\in \mathbb{C}^{N_{e} \times N_{\rm{AA}}}$ is the channel matrix from the $i$th TAS to Eve.
$\textbf{n}_b\sim\mathcal{CN}(0,\sigma_b^2\textbf{I}_{N_b})$ and $\textbf{n}_e\sim\mathcal{CN}(0,\sigma_e^2\textbf{I}_{N_e})$ denote the complex additive white Gaussian noise (AWGN) vectors at Bob and Eve, respectively.
Thus, the received signals observed at Bob and Eve after decoding processing can be respectively formulated as follows
\begin{align} \label{yb}  \nonumber
\textbf{y}_b=&\sqrt{\beta P}\textbf{W}_b^H\textbf{HE}_i\textbf{F}_{\rm{RF}}\textbf{F}_{\rm{BB}}b_j \\ +&\sqrt{(1-\beta)P}\textbf{W}_b^H\textbf{HT}\textbf{F}_{\rm{RF}}\textbf{T}_{\rm{BB}}\textbf{n}+\textbf{W}_b^H\textbf{n}_b,
\end{align}
\begin{align} \label{ye}  \nonumber
\textbf{y}_e=&\sqrt{\beta P}\textbf{W}_e^H\textbf{GE}_i\textbf{F}_{\rm{RF}}\textbf{F}_{\rm{BB}}b_j \\ +&\sqrt{(1-\beta)P}\textbf{W}_e^H\textbf{GT}\textbf{F}_{\rm{RF}}\textbf{T}_{\rm{BB}}\textbf{n}+\textbf{W}_e^H\textbf{n}_e,
\end{align}
where $\textbf{W}_b\!\in\! \mathbb{C}^{N_b \times N_{\rm{RF}}}$ is the combiner of Bob. It can be expanded as $\textbf{W}_b\!=\![\textbf{w}_1^b,\textbf{w}_2^b,\ldots,\textbf{w}_{N_{\rm{RF}}}^b]$, and $\textbf{w}_i^b\in \mathbb{C}^{N_{b} \times 1},i\!=\!1,2,\ldots,N_{\rm{RF}}$.
$\textbf{W}_e\!\in\! \mathbb{C}^{N_e \times N_{\rm{RF}}}$ is the combiner of Eve, and $\textbf{W}_e\!=\![\textbf{w}_1^e,\textbf{w}_2^e,\ldots,\textbf{w}_{N_{\rm{RF}}}^e]$, where $\textbf{w}_i^e\!\in\! \mathbb{C}^{N_{e} \times 1},i\!=\!1,2,\ldots,N_{\rm{RF}}$.

The maximum-likelihood (ML) detector at Bob is given by
\begin{align} \label{ML}
[\hat{i},\hat{j}]=&\mathop{\textrm{arg} \ \textrm{min}}\limits_{i\in I_{\rm{ASS}},j\in \mathbb{M}} \ \|\textbf{y}_b-\sqrt{\beta P}\textbf{W}_b^H\textbf{HE}_i\textbf{F}_{\rm{RF}}\textbf{F}_{\rm{BB}}b_j\|^2,
\end{align}
%=&\mathop{\textrm{arg} \ \textrm{min}}\limits_{i\in I_{\rm{AAS}},j\in\mathbb{M}} \ \|\textbf{y}_b-\sqrt{\beta P}\textbf{W}_b^H\textbf{h}_i\textbf{f}_ip_ib_j\|^2

The average SR is given as
\begin{align}
\overline{R_s}=\mathbb{E}_{\textbf{H},\textbf{G}}[I(\textbf{x};\textbf{y}_b)-I(\textbf{x};\textbf{y}_e)]^+,
\end{align}
where
\begin{align}  \nonumber
I(\textbf{x};\textbf{y}_b')&=\textrm{log}_2N_tM - \frac{1}{N_tM}\times \\ & \sum \limits_{i = 1}^{N_tM} {\mathbb{E}_{\textbf{n}_b'} \left\{ \textrm{log}_2 \sum \limits_{j=1} ^{N_tM} \textrm{exp} \left(-f_{b,i,j}+\|\textbf{n}_b'\|^2 \right) \right\}},
\end{align}
\begin{align}  \nonumber
I(\textbf{x};\textbf{y}_e')&=\textrm{log}_2N_tM - \frac{1}{N_tM}\times \\ & \sum \limits_{i = 1}^{N_tM} {\mathbb{E}_{\textbf{n}_e'} \left\{ \textrm{log}_2 \sum \limits_{j=1} ^{N_tM} \textrm{exp} \left(-f_{e,m,k}+\|\textbf{n}_e'\|^2 \right) \right\}},
\end{align}
where $\textbf{y}'_b\!=\!\boldsymbol{\Omega}_{\rm{B}}^{-1/2}\textbf{y}_b$, $\textbf{y}'_e\!=\!\boldsymbol{\Omega}_{\rm{E}}^{-1/2}\textbf{y}_e$, $\textbf{n}'_b\!=\!\boldsymbol{\Omega}_{\rm{B}}^{-1/2}\textbf{W}_b^H\textbf{n}_b$, and $\textbf{n}'_e\!=\!\boldsymbol{\Omega}_{\rm{E}}^{-1/2}\textbf{W}_e^H\textbf{n}_e$,
\begin{align}
f_{b,i,j}=\|\sqrt{\beta P}\boldsymbol{\Omega}_{\rm{B}}^{-1/2}\textbf{W}_b^H\textbf{H}\textbf{d}_{ij}+\textbf{n}'_b\|^2,
\end{align}
\begin{align}
f_{e,m,k}=\|\sqrt{\beta P}\boldsymbol{\Omega}_{\rm{E}}^{-1/2}\textbf{W}_e^H\textbf{G}\textbf{d}_{mk}+\textbf{n}'_e\|^2,
\end{align}
$\textbf{d}_{ij}\!=\!\textbf{x}_i\!-\!\textbf{x}_j$ and $\textbf{d}_{mk}\!=\!\textbf{x}_m\!-\!\textbf{x}_k$,
where $\textbf{x}_i$, $\textbf{x}_j$, $\textbf{x}_m$, and $\textbf{x}_k$ are the possible transmit vectors. $\boldsymbol{\Omega}_{\rm{B}}$ and $\boldsymbol{\Omega}_{\rm{E}}$ are the covariance matrices of interference plus noise of Bob and Eve respectively, where
\begin{align}
\boldsymbol{\Omega}_{\rm{B}}=(1-\beta)P\textbf{C}_b + \sigma_b^2\textbf{I}_{N_{\rm{RF}}},
\end{align}
\begin{align}
\boldsymbol{\Omega}_{\rm{E}}=(1-\beta)P\textbf{C}_e +  \sigma_e^2\textbf{I}_{N_{\rm{RF}}},
\end{align}
with
\begin{align}
\textbf{C}_b\!=\!\textbf{W}_b^H\textbf{HT}\textbf{F}_{\rm{RF}}\textbf{T}_{\rm{BB}}\textbf{T}_{\rm{BB}}^H\textbf{F}_{\rm{RF}}^H\textbf{T}^H\textbf{H}^H\textbf{W}_b,
\end{align}
\begin{align}
\textbf{C}_e\!=\!\textbf{W}_e^H\textbf{GT}\textbf{F}_{\rm{RF}}\textbf{T}_{\rm{BB}}\textbf{T}_{\rm{BB}}^H\textbf{F}_{\rm{RF}}^H\textbf{T}^H\textbf{G}^H\textbf{W}_e,
\end{align}
respectively.
Since $\boldsymbol{\Omega}_{\rm{B}}$ and $\boldsymbol{\Omega}_{\rm{E}}$ are adopted to whiten colored noise into an white noise, they don't change the mutual information, that is, $I(\textbf{x};\textbf{y}_b)\!=\!I(\textbf{x};\textbf{y}'_b)$ and $I(\textbf{x};\textbf{y}_e)\!=\!I(\textbf{x};\textbf{y}'_e)$.

In addition, to facilitate the hybrid precoding design, $\textbf{P}_{\rm{AN}}$ is chosen to be the projection matrix in the null space of channel $\textbf{H}$, satisfying $\textbf{W}_b^H\textbf{HT}\textbf{F}_{\rm{RF}}\textbf{T}_{\rm{BB}}=\textbf{W}_b^H\textbf{HTP}_{\rm{AN}}=\textbf{0}$. Therefore,
\begin{align}
\textbf{P}_{\rm{AN}}=\frac{1}{\mu}\left[\textbf{I}-\textbf{H}_\textbf{T}'^H(\textbf{H}_\textbf{T}'\textbf{H}_\textbf{T}'^H)^{\dagger}\textbf{H}_\textbf{T}'\right],
\end{align}
where $\textbf{H}_\textbf{T}\!=\![\textbf{h}_{l_1},\textbf{h}_{l_2},\cdots,\textbf{h}_{l_{N_t}}]\!\in\! \mathbb{C}^{N_b \times N_{\rm{AA}}N_t}$ is the channel matrix corresponding to the selected subchannels $I_{\rm{ASS}}\!=\!\{l_i\}_{i=1}^{N_t}$. Due to the relation $\textbf{P}_{\rm{AN}}\!=\!\textbf{F}_{\rm{RF}}\textbf{T}_{\rm{BB}}$,
 once $\textbf{F}_{\rm{RF}}$ is determined, so $\textbf{T}_{\rm{BB}}$ can be obtained readily.

The analog precoding corresponding to the selected channel is expressed as
\begin{equation} \label{AT}
\textbf{F}_{\rm{RF},\textbf{T}}={
\left[
\begin{array}{cccc}
    \textbf{f}_{l_1} & \textbf{0} & \cdots & \textbf{0} \\
    \textbf{0} & \textbf{f}_{l_2} & \cdots & \textbf{0} \\
    \vdots & \vdots & \ddots & \vdots \\
    \textbf{0} & \textbf{0} & \cdots & \textbf{f}_{l_{N_t}}
\end{array}
\right]_{N_{\rm{AA}}N_t \times N_t}}.
\end{equation}
And the digital precoding matrix corresponding to the selected channel is $\textbf{T}_{\rm{BB},\textbf{T}}\!\in\! \mathbb{C}^{N_t \times N_t}$ and  satisfies $\textbf{T}_{\rm{BB},\textbf{T}}\!=\!\textbf{ST}_{\rm{BB}}$, where $\textbf{S}$ is the selection matrix, and the $l_i$-th column of $\textbf{S}$ is the $i$-th column of $\textbf{I}_{N_t}$.

Therefore, $\textbf{W}_b^H\textbf{HT}\textbf{F}_{\rm{RF}}\textbf{T}_{\rm{BB}}\!=\!\textbf{W}_b^H\textbf{H}_\textbf{T}\textbf{F}_{\rm{RF},\textbf{T}}\textbf{T}_{\rm{BB},\textbf{T}}\!=\!\textbf{0}$. Let us define $\textbf{H}'\!=\!\textbf{W}_b^H\textbf{H}_\textbf{T}\textbf{F}_{\rm{RF},\textbf{T}}$, $\textbf{T}_{\rm{BB},\textbf{T}}$ can be given by
\begin{align} \label{Tbb}
\textbf{T}_{\rm{BB},\textbf{T}}=\frac{1}{\mu}[\textbf{I}-\textbf{H}'^H(\textbf{H}'\textbf{H}'^H)^{\dagger}\textbf{H}'],
\end{align}
where $\mu$ is the normalized factor. According to the power constraint at transmit side, and considering $\|\textbf{T}\textbf{F}_{\rm{RF}}\textbf{T}_{\rm{BB}}\|_F^2\!=\!\|\textbf{F}_{\rm{RF},\textbf{T}}\textbf{T}_{\rm{BB},\textbf{T}}\|_F^2\!=\!\|\textbf{T}_{\rm{BB},\textbf{T}}\|_F^2\!=\!1$, then we have $\|\textbf{T}_{\rm{BB},\textbf{T}}\|_F^2\!=\!1$ and  $\mu\!=\!\|\textbf{I}\!-\!\textbf{H}'^H(\textbf{H}'\textbf{H}'^H)^{\dagger}\textbf{H}'\|_\textrm{F}$. Therefore, we obtain the digital precoder $\textbf{T}_{\rm{BB}}\!=\!\textbf{S}^T\textbf{T}_{\rm{BB},\textbf{T}}$ of AN.
$\textbf{A}^\dagger$ is the Moore-Penrose pseudo inverse of $\textbf{A}$.

\section{Proposed  Hybrid Precoding methods}
For hybrid SM systems, the design of hybrid precoder at Alice and combiner at Bob is very important to improve the system performance. In this section, the approximate expression of SR is derived by using the definition of cut-off rate in \cite{Aghdam2017Joint}. Then, two hybrid precoders, called Max-ASR-ADMM and Max-ASR-GA, are proposed to enhance the security of SM systems. Additionally, the SDR-AltMin Algorithm in \cite{Yu2016Alternating} is also extended to be suitable for hybrid SM and used as a performance reference.

\subsection{Approximate expression of SR}
In accordance with the definition of the cut-off rate for traditional MIMO systems in \cite{Aghdam2017Joint}, we have the cut-off rate for Bob
\begin{align} \label{Ib}
\!\!I_0^{B}\!\!= \!\!-\!\log_2\!\!\sum \limits_{i=1}^{N_tM}\!\sum \limits_{j=1}^{N_tM}\!\frac{1}{(N_tM)^2}\!\!\int\! \!p(\textbf{y}_b|\textbf{x}_i)^{1/2}  p(\textbf{y}_b|\textbf{x}_j)^{1/2} d\textbf{y}_b.\!\!
\end{align}
For a given  channel $\textbf{H}$, assuming the receive signal $\textbf{y}_b$ is a complex Gaussian distribution, and the corresponding conditional probability is
\begin{align} \label{pB}
p(\textbf{y}'_b|\textbf{x}_i)=\frac{1}{(\pi\sigma_b^2)^{N_r}}\exp\left( \|(\textbf{y}'_b-\sqrt{\beta P}\textbf{W}'^H_b\textbf{H}\textbf{x}_i)\|^2 \right),
\end{align}
where $\textbf{y}'_b\!=\!\boldsymbol{\Omega}_{\rm{B}}^{-1/2}\textbf{y}_b$ and $\textbf{W}_b'^H\!=\!\boldsymbol{\Omega}_{\rm{B}}^{-1/2}\textbf{W}_b^H$.
Making use of (\ref{pB}), we have
\begin{align} \nonumber
\!I_0^{B}\!=&2\textrm{log}_2N_tM- \\
&\!\log_2\!\sum \limits_{i=1}^{N_tM}\!\sum \limits_{j=1}^{N_tM}\!\exp\!\left(\! \frac{\!-\!\beta P\textbf{d}_{ij}^H\textbf{H}^H\textbf{W}_b\boldsymbol{\Omega}_{\rm{B}}^{-1}\textbf{W}_b^H\textbf{H}\textbf{d}_{ij}}{4} \!\right)\!,
\end{align}
which can be derived  similar to Appendix A in \cite{Aghdam2017Joint} with a slight modification. Similarly, the cut-off rate $I_0^E$ for Eve is given by
\begin{align} \nonumber
\!I_0^{E}\!=&2\textrm{log}_2N_tM- \\
&\!\log_2\!\sum \limits_{i=1}^{N_tM}\!\sum \limits_{j=1}^{N_tM}\!\exp\!\left( \! \frac{\!-\!\beta P\textbf{d}_{ij}^H\textbf{G}^H\textbf{W}_e\boldsymbol{\Omega}_{\rm{E}}^{-1}\textbf{W}_e^H\textbf{G}\textbf{d}_{ij}}{4} \!\right)\!.
\end{align}
Since the ASR can be expressed as $R_s^{a}=I_0^{B}-I_0^{E}$, the optimization problem of maximizing ASR can be casted as
\begin{align}\label{OSR}
&\max ~~  R_s^{a}(\textbf{P})\\ \nonumber
&\textrm{subject} \ \textrm{to} \ \|\textbf{P}\|^2=N_{\rm{RF}},
\end{align}
where $\textbf{P}=\textbf{F}_{\rm{RF}}\textbf{F}_{\rm{BB}}$ is the total precoding matrix, and
\begin{align} \label{Rsa}
R_s^{a}(\textbf{P})= \textrm{log}_2\kappa_E(\textbf{P})- \textrm{log}_2\kappa_B(\textbf{P}),
\end{align}
where
\begin{align} \label{kB}
&\!\kappa_B(\textbf{P})\!=\!\sum \limits_{i=1}^{N_tM}\!\sum \limits_{j=1}^{N_tM}\!\exp\!\left(\! \frac{-\beta P\textbf{d}_{ij}^H\textbf{H}^H\textbf{W}_b\boldsymbol{\Omega}_{\rm{B}}^{-1}\textbf{W}_b^H\textbf{H}\textbf{d}_{ij}}{4} \!\right), \! \\ \label{kE}
&\!\kappa_E(\textbf{P})\!=\! \sum \limits_{i=1}^{N_tM}\!\sum \limits_{j=1}^{N_tM}\!\exp\!\left(\! \frac{-\beta P\textbf{d}_{ij}^H\textbf{G}^H\textbf{W}_e\boldsymbol{\Omega}_{\rm{E}}^{-1}\textbf{W}_e^H\textbf{G}\textbf{d}_{ij}}{4} \!\right),\!
\end{align}
where $\textbf{d}_{ij}\!=\!\textbf{x}_i(\textbf{P})\!-\!\textbf{x}_j(\textbf{P})$, where  $\textbf{x}_i(\textbf{P})$ and $\textbf{x}_j(\textbf{P})$ are randomly chosen from the set of all possible combinations of $\textbf{x}$ shown in (\ref{x-def}).
%$=\textbf{E}_i\textbf{F}_{\rm{RF}}\textbf{F}_{\rm{BB}}b_j=\textbf{E}_i\textbf{P}b_j$, we can get $\textbf{d}_{ij}(\textbf{P})=\textbf{x}_i(\textbf{P})-\textbf{x}_j(\textbf{P})$ and $\textbf{d}_{mk}(\textbf{P})=\textbf{x}_m(\textbf{P})-\textbf{x}_k(\textbf{P})$.

Due to the non-convexity of the optimization problem in (\ref{OSR}), it is very difficult to obtain its global optimal solution directly. Below, two methods are proposed to address the optimization problem in (\ref{OSR}).

\subsection{Proposed Max-ASR-ADMM}
Observing the expression of $\textbf{P}$ in (\ref{P-def}), we find only one precoding $\textbf{p}_i$ corresponding to the active TAS of the hybrid precoding matrix $\textbf{P}$ is non-zero in the actual communication process, so $\textbf{d}_{ij}(\textbf{P})$ is at most two subarrays with non-zero values, which means that $\textbf{P}$ is sparse in the actual process of maximizing the SR. For such a high-dimensional but sparse optimization problem, the data of the optimization problem can be segmented and then transformed into a general form consensus problem, which can be solved by using ADMM. With that in mind, the expression  in (\ref{kB}) can be further simplified as follows
\begin{align}\label{b1}
\kappa_B
\!=\!&\sum \limits_{\substack{m,m'\\ \in I_{\rm{ASS}}}}\!\sum \limits_{k,k'=1}^{M}\!\exp\left( \frac{-\beta P\textbf{q}_{m,m'}^H\textbf{B}_{m,m'}^{k,k'}\textbf{q}_{m,m'}}{4} \right)\\ \label{b2}
\!=\!&\sum \limits_{\substack{m,m'\\ \in I_{\rm{ASS}}}}\!\sum \limits_{k,k'=1}^{M}\!\exp\left( \frac{-\beta P\textrm{Tr}(\textbf{q}_{m,m'}^H\textbf{B}_{m,m'}^{k,k'}\textbf{q}_{m,m'})}{4} \right)\\ \label{b3}
\!=\!&\sum \limits_{\substack{m,m'\\ \in I_{\rm{ASS}}}}\!\sum \limits_{k,k'=1}^{M}\!\exp\left( \frac{-\beta P\textrm{Tr}(\textbf{q}_{m,m'}\textbf{q}_{m,m'}^H\textbf{B}_{m,m'}^{k,k'})}{4} \right),
\end{align}
where
\begin{align}
\textbf{q}_{m,m'}=
\left\{
             \begin{array}{cc}
             \begin{bmatrix} \textbf{p}_m \\ \textbf{p}_{m'}\end{bmatrix}, & m\neq m' \\
             \textbf{p}_m, & m=m'\\
             \end{array}
\right.
\end{align}
is the precoders corresponding to those activated TASs.
\begin{align} \label{B_mm^kk}
\!\textbf{B}_{m,m'}^{k,k'}\!=\!
\left\{\!\!
             \begin{array}{ll}
             \!\textbf{J}_{k,k'}^H\textbf{H}_{m,m'}^H\!\textbf{W}_b\boldsymbol{\Omega}_{\rm{B}}^{-1}\textbf{W}_b^H\textbf{H}_{m,m'}\!\textbf{J}_{k,k'}, \!\!\!&\! m\neq\! m' \\
             \!\textbf{J}_{k,k'}^H\textbf{h}_{m}^H\textbf{W}_b\boldsymbol{\Omega}_{\rm{B}}^{-1}\textbf{W}_b^H\textbf{h}_{m}\textbf{J}_{k,k'}, \!\!\!& \!m\!=\!m'\!\\
             \end{array}
\!\right.\!
\end{align}
where
\begin{align}
\textbf{H}_{m,m'}=\left[\textbf{h}_{m} \ \textbf{h}_{m'}\right]
\end{align}
is the channel matrix corresponding to those activated TASs.
\begin{align}
{{\mathbf{J}}_{k,k'}} = \left\{ {\begin{array}{*{20}{l}}
{{\mathbf{b}} \otimes {{\mathbf{I}}_{{N_{{\text{AA}}}}}},}&{m \ne m'} \\
{({b_k} - {b_{k'}}){{\mathbf{I}}_{{N_{{\text{AA}}}}}},}&{m = m'}
\end{array}} \right.
\end{align}
is the sending symbol, where
\begin{align}
\textbf{b}=
\left[
             \begin{array}{cc}
             b_k & 0 \\
             0 & -b_{k'}\\
             \end{array}
\right],
\end{align}
$\otimes$ is the Kronecker product of two matrices.
The above simplification addresses the problem that it is hard to solve the precoding matrix $\textbf{P}$  due to its high-dimensional and sparse characteristic. The derivation from (\ref{b1}) to (\ref{b3}) is achieved by utilizing the trace property, i.e., $\textrm{tr}(\textbf{AB}) = \textrm{tr}(\textbf{BA})$ for matrices $\textbf{A}$ and $\textbf{B}$.

Similarly, for Eve, $\kappa_E$ can be simplified to
\begin{align}
\kappa_E\!=\!\sum \limits_{\substack{m,m'\\ \in I_{\rm{ASS}}}}\!\sum \limits_{k,k'=1}^{M}\!\exp\!\left(\! \frac{-\beta P\textrm{Tr}(\textbf{q}_{m,m'}\textbf{q}_{m,m'}^H\textbf{E}_{m,m'}^{k,k'})}{4} \!\right),
\end{align}
where
\begin{align}\label{E_mm^kk}
\!\textbf{E}_{m,m'}^{k,k'}\!=\!
\left\{\!
             \begin{array}{ll}
             \!\textbf{J}_{k,k'}^H\textbf{G}_{m,m'}^H\!\textbf{W}_e\boldsymbol{\Omega}_{\rm{E}}^{-1}\textbf{W}_e^H\textbf{G}_{m,m'}\!\textbf{J}_{k,k'},\!\!\! & \!m\!\neq\! m' \\
             \!\textbf{J}_{k,k'}^H\textbf{g}_{m}^H\textbf{W}_e\boldsymbol{\Omega}_{\rm{E}}^{-1}\textbf{W}_e^H\textbf{g}_{m}\textbf{J}_{k,k'},\!\!\! & \!m\!=\!m'\!\\
             \end{array}
\!\right.\!
\end{align}
\begin{align}
\textbf{G}_{m,m'}=\left[\textbf{g}_{m} \ \textbf{g}_{m'}\right]
\end{align}
is the channel matrix corresponding to those activated TASs.

By using the Jensen's inequality, the lower bound of  $R_s^{a}(\textbf{P})$  is
\begin{align} \nonumber
&R_s^{a'}(\textbf{P})\\ \nonumber
&\!=\!\log_2\!\exp\!\sum \limits_{\substack{m,m'\\ \in I_{\rm{ASS}}}}\!\sum \limits_{k,k'=1}^{M}\!\left(\! \frac{-\beta P\textrm{Tr}(\textbf{q}_{m,m'}\textbf{q}_{m,m'}^H\textbf{E}_{m,m'}^{k,k'})}{4} \!\right)\!\\
&\!-\!\log_2\!\exp\!\sum \limits_{\substack{m,m'\\ \in I_{\rm{ASS}}}}\!\sum \limits_{k,k'=1}^{M}\!\left(\! \frac{-\beta P\textrm{Tr}(\textbf{q}_{m,m'}\textbf{q}_{m,m'}^H\textbf{B}_{m,m'}^{k,k'})}{4}\! \right)\! \\ \nonumber
&\!=\!\log_2e\!\cdot\!\left[\!\sum \limits_{\substack{m,m'\\ \in I_{\rm{ASS}}}}\!\sum \limits_{k,k'=1}^{M}\!\left(\! \frac{-\beta P\textrm{Tr}(\textbf{q}_{m,m'}\textbf{q}_{m,m'}^H\textbf{E}_{m,m'}^{k,k'})}{4} \! \right)\!\right.\!\\
&\qquad ~~~-\left.\!\sum \limits_{\substack{m,m'\\ \in I_{\rm{ASS}}}}\!\sum \limits_{k,k'=1}^{M}\!\left( \!\frac{-\beta P\textrm{Tr}(\textbf{q}_{m,m'}\textbf{q}_{m,m'}^H\textbf{B}_{m,m'}^{k,k'})}{4} \! \right)\!\right]\!,\!
\end{align}
which yields the following optimization problem
\begin{align}\label{OSR-B}
&\max ~~  R_s^{a'}(\textbf{P})\\ \nonumber
&\textrm{subject} \ \textrm{to} \ \|\textbf{P}\|^2=N_{\rm{RF}}.
\end{align}
Let us define the following function
\begin{align} \nonumber \label{f}
f_{m,m'}&(\textbf{q}_{m,m'})\!=\!\left[\!\sum \limits_{k,k'=1}^{M}\!\left(\! \frac{-\beta P\textrm{Tr}(\textbf{q}_{m,m'}\textbf{q}_{m,m'}^H\textbf{B}_{m,m'}^{k,k'})}{4} \!\right)\!\right.\!\\
&\!-\!\left.\!\sum \limits_{k,k'=1}^{M}\!\left(\! \frac{-\beta P\textrm{Tr}(\textbf{q}_{m,m'}\textbf{q}_{m,m'}^H\textbf{E}_{m,m'}^{k,k'})}{4}\! \right)\!\right]\!\cdot\!\log_2e,
\end{align}
then, the optimization problem in  (\ref{OSR-B}) reduces to a general form consensus problem
\begin{align} \label{OSR-B-ADMM} \nonumber
\mathop{\textrm{minimize}} \ & \sum_{\substack{m,m'\in I_{\rm{ASS}}}} f_{m,m'}(\textbf{q}_{m,m'})  \\ \nonumber
\textrm{subject} \ \textrm{to} \ & \|\textbf{P}\|^2=N_{\rm{RF}}\\
&\textbf{q}_{m,m'}-\textbf{p}_{m,m'}=0, \  m,m' \in I_{\rm{ASS}},
\end{align}
where $\textbf{p}_{m,m'}$ are functions of $\textbf{P}$, $\textbf{p}_{m,m'}=\mathbf{L}_{m,m'}\textbf{P}$, and $\mathbf{L}_{m,m'}\in \mathbb{R}^{2N_{\rm{AA}} \times N_{\rm{AA}}N_{\rm{RF}}}$ assign the precoding of each TAS to the corresponding position in the precoding matrix $\textbf{P}$.
$(\mathbf{L}_{m,m'})_{1:N_{\rm{AA}},l:l+N_{\rm{AA}}}=\textbf{I}_{N_{\rm{AA}}}$, $l=(m-1)N_{\rm{AA}}+1$,  $(\mathbf{L}_{m,m'})_{N_{\rm{AA}}+1:2N_{\rm{AA}},l':l'+N_{\rm{AA}}}=\textbf{I}_{N_{\rm{AA}}}$,  $l'=(m'-1)N_{\rm{AA}}+1$, and the rest of $\mathbf{L}_{m,m'}$ is zero.

The above general form consensus problem can be solved by ADMM \cite{boyd2011distributed}.
Let us introduce a new matrix variable $\textbf{Q}_{m,m'}=\textbf{q}_{m,m'}\textbf{q}_{m,m'}^H$, then we obtain the equivalence of (\ref{OSR-B-ADMM}) as follows
\begin{align} \nonumber
f_{m,m'}&(\textbf{Q}_{m,m'})=\frac{-\beta P\log_2e}{4}\cdot \\
&\sum \limits_{k,k'=1}^{M}\left[ \textrm{Tr}(\textbf{Q}_{m,m'}\textbf{B}_{m,m'}^{k,k'})-\textrm{Tr}(\textbf{Q}_{m,m'}\textbf{E}_{m,m'}^{k,k'}) \right].
\end{align}
Taking the above expression to replace the loss function in (\ref{OSR-B-ADMM})  and introduce dual variables ${{\mathbf{Y}}_{m,m'}}$, we have the alternating iterative process of ADMM including three main expressions
\begin{align} \nonumber \label{optadmm}
\mathop{\textrm{minimize}}\limits_{\substack{\textbf{Q}_{m,m'}}} \ \ &\Big(f_{m,m'}(\textbf{Q}_{m,m'})+ \\ \nonumber
&\left.{\textbf{Y}_{m,m'}^t}^H\textbf{Q}_{m,m'}+\left(\frac{\rho}{2}\right)\|\textbf{Q}_{m,m'}-\textbf{P}_{m,m'}^t\|_2^2\right)\\
\textrm{subject} \ \textrm{to}\ &\textrm{Tr}(\textbf{Q}_{m,m'})=
\left\{
             \begin{array}{lr}
             2,\  m\neq m' \\
             1,\  m=m'\\
             \end{array}
\right.\\ \nonumber
&\textbf{Q}_{m,m'}\succeq 0\\ \nonumber
&{\textbf{Q}_{m,m'}} - \textbf{P}_{m,m'}^t = {\textbf{0}},
\end{align}
\begin{align} \label{P-ADMM}
\textbf{P}=\frac{\sqrt{N_{\rm{RF}}}\cdot\sum \limits_{\substack{m,m' \in I_{\rm{ASS}}}}\mathbf{L}_{m,m'}^T\textbf{q}_{m,m'}^{t+1}}{\left\|\sum \limits_{\substack{m,m' \in I_{\rm{ASS}}}}\mathbf{L}_{m,m'}^T\textbf{q}_{m,m'}^{t+1}\right\|_2},
\end{align}
and
\begin{align} \label{Y}
\textbf{Y}_{m,m'}^{t+1}=\textbf{Y}_{m,m'}^t+\rho(\textbf{Q}_{m,m'}^{t+1}-\textbf{P}_{m,m'}^{t+1}),
\end{align}
where $\textbf{P}_{m,m'}=\mathbf{p}_{m,m'}\mathbf{p}_{m,m'}^H$.
The third term of (\ref{optadmm}) $\left( {{\rho  \mathord{\left/ {\vphantom {\rho  2}} \right. \kern-\nulldelimiterspace} 2}} \right)$ $\left\| {{{\mathbf{Q}}_{m,m'}} - {\mathbf{P}}_{m,m'}^t} \right\|_2^2$ is the penalty term.  Variable $\textbf{P}$ in (\ref{P-ADMM})   deals with the global variable $\textbf{P}$ and is also called the central collector or the fusion center. By introducing matrix $\textbf{Q}_{m,m'}$, the optimization problem in (\ref{optadmm}) is converted into a convex semidefinite programming (SDP) problem, which can be solved conveniently by using convex optimization toolbox CVX.
Then, let $\textbf{f}_i=\frac{1}{\sqrt{N_{\rm{AA}}}}e^{j\textrm{angle}(\textbf{p}_i)}$ as the analog precoder of the $i$-th TAS and $p_i=\|\textbf{p}_i\|$ as the digital precoder of the $i$-th TAS. Now, we complete the design of hybrid precoding.
Therefore, a step-by-step summary is provided as follows: \textbf{Algorithm \ref{alg:1}}.

%\textbf{Proposed Max-ASR-ADMM Hybrid Precoding Scheme}:\\
%\textbf{Step 1}: Initializes the precoding matrix $\textbf{P}_0=\textbf{V}(:,1)$, so the $\textbf{P}_{m,m'}^0$ can be determined for all $m,m' \in I_{\rm{ASS}}$, and initializes $\textbf{Q}_{m,m'}^0$, $\textbf{Y}_{m,m'}^0$, $\rho=0.5$ and $t=0$.\\
%\textbf{Step 2}: For all $m,m' \in I_{\rm{ASS}}$, alternate iteration between (\ref{optadmm}) and (\ref{Y}), therefore, we obtain $\textbf{Q}_{m,m'}^{t+1}$ and $\textbf{Y}_{m,m'}^{t+1}$ , and  $\textbf{q}_{m,m'}^{t+1}$.\\
%\textbf{Step 3}: Substitute $\textbf{q}_{m,m'}^{t+1}$ into the fusion center  (\ref{P-ADMM}), and $\textbf{P}_{t+1}$ can be obtained.\\
%\textbf{Step 4}: Judge whether the termination condition $\left\|\textbf{P}_{t+1}-\textbf{P}_{t}\right\|_2\!<\!0.01$ is satisfied. If so, the iteration will terminate and output the hybrid precoding matrix $\textbf{P}=\textbf{P}_{t+1}$. Otherwise, let $t=t+1$, and return to Step 2 for iteration.\\
%\textbf{Step 5}: Let $\textbf{f}_i=\frac{1}{\sqrt{N_{\rm{AA}}}}e^{j\textrm{angle}(\textbf{p}_i)}$ as the analog precoder of the $i$-th TAS and $p_i=\|\textbf{p}_i\|$ as the digital precoder of the $i$-th TAS. Now, we complete the design of hybrid precoding.

\begin{algorithm}
	\renewcommand{\algorithmicrequire}{\textbf{Input:}}
	\renewcommand{\algorithmicensure}{\textbf{Output:}}
	\caption{Max-ASR-ADMM hybrid precoding scheme}
	\label{alg:1}
	\begin{algorithmic}[1]
		\REQUIRE the channel matrix $\textbf{H}$ and $\textbf{G}$, the M-ary constellation, and TASS matrix $\textbf{T}$
		\ENSURE $\textbf{P}$, $\textbf{F}_{\rm{RF}}$, $\textbf{F}_{\rm{BB}}$
		\STATE Compute $\textbf{B}_{m,m'}^{k,k'}$ and $\textbf{E}_{m,m'}^{k,k'}$ for each $m,m' \in I_{\rm{ASS}}$ according to (\ref{B_mm^kk}) and (\ref{E_mm^kk})
        \STATE Initialize $\rho=0.5$ and $t=0$
		\STATE Initialize the precoding matrix $\textbf{P}_0=\textbf{V}(:,1)$
		\STATE Determine $\textbf{P}_{m,m'}^0$ for each $m,m' \in I_{\rm{ASS}}$ according to $\textbf{P}_0$
		\STATE Initializes $\textbf{Q}_{m,m'}^0$, $\textbf{Y}_{m,m'}^0$ similar to $\textbf{P}_{m,m'}^0$
        \REPEAT
		\FORALL{$m,m' \in I_{\rm{ASS}}$}
        \STATE Let $t=t+1$
		\STATE Update $\textbf{Q}_{m,m'}^{t}$ according to (\ref{optadmm})
		\STATE Update $\textbf{Y}_{m,m'}^{t}$ according to $\textbf{Q}_{m,m'}^{t}$ and (\ref{Y})
		\STATE Update $\mathbf{q}_{m,m'}^{t}$ according to $\textbf{Q}_{m,m'}=\mathbf{q}_{m,m'}\mathbf{q}_{m,m'}^H$
		\ENDFOR
        \STATE Update $\textbf{P}_{t}$ based on $\mathbf{q}_{m,m'}^{t+1}$ according to (\ref{P-ADMM})
        \UNTIL{$\left\|\textbf{P}_{t}-\textbf{P}_{t-1}\right\|_2\!<\!0.01$}
        \STATE Compute $\textbf{f}_i$ and $p_i$ as the $i$-th TAS analog and digital precoders
        \STATE Substitute $\textbf{f}_i$ and $p_i$ into (\ref{F_rf}) and (\ref{F_bb})
        \STATE \textbf{return} $\textbf{P}=\textbf{P}_{t}$, $\textbf{F}_{\rm{RF}}$, $\textbf{F}_{\rm{BB}}$
	\end{algorithmic}
\end{algorithm}

\subsection{Proposed Max-ASR-GA}
In the previous section, the Max-ASR-ADMM algorithm is presented to optimize the secure hybrid precoding matrices. For the comparison of  the secrecy performance and  to offer a new solution to this non-convex optimization problem, we propose another secure hybrid precoding scheme,  namely Max-ASR-GA, in what follows. To maximize $R_s^a(\textbf{P})$, the Max-ASR-GA method can be employed to directly optimize the precoding matrix $\textbf{P}$. The gradient vector of $R_s^a(\textbf{P})$  with respect to  $\textbf{P}$ can be derived as
\begin{equation}
\begin{aligned} \label{gradient}
&\!\nabla_{\textbf{P}} R_s^{a}(\textbf{P})=\sum \limits_{i=1}^{N_tM}\sum \limits_{j=1}^{N_tM} \frac{\partial R_s^{a}(\textbf{P})}{\partial \textbf{d}_{ij}}\frac{\partial \textbf{d}_{ij}}{\partial \textbf{P}}=\frac{-\beta P}{4\ln2}\cdot\\
&\!\!\left\{\!\!\frac{1}{\kappa_E}\!\sum \limits_{i=1}^{N_tM}\!\sum \limits_{j=1}^{N_tM}\!\exp\!\left[\! \frac{-\beta P\textbf{d}_{ij}^H\textbf{H}^H\textbf{W}_e\boldsymbol{\Omega}_{\rm{E}}^{-1}\textbf{W}_e^H\textbf{H}\textbf{d}_{ij}}{4} \!\right]\!\cdot\!\bm{\chi}_{\rm{E}}\right.\\
&\!\!\!\left.\!-\frac{1}{\kappa_B}\!\sum \limits_{i=1}^{N_tM}\!\sum \limits_{j=1}^{N_tM}\!\exp\!\left[\! \frac{-\beta P\textbf{d}_{ij}^H\textbf{G}^H\textbf{W}_b\boldsymbol{\Omega}_{\rm{B}}^{-1}\textbf{W}_b^H\textbf{G}\textbf{d}_{ij}}{4} \!\right]\!\cdot\!\bm{\chi}_{\rm{B}}\!\!\right\}\!
\end{aligned}
\end{equation}
where
\begin{align} \label{X_B}
\bm{\chi}_{\rm{B}}=\textbf{D}_{ij}\left[\textbf{A}_{\rm{H}}+\textbf{A}_{\rm{H}}^H\right]\textbf{d}_{ij},
\end{align}
and
\begin{align} \label{X_E}
\bm{\chi}_{\rm{E}}=\textbf{D}_{ij}\left[\textbf{A}_{\rm{G}}+\textbf{A}_{\rm{G}}^H\right]\textbf{d}_{ij}.
\end{align}
Let $\textbf{X}=\textbf{E}_ib_j$, then $\textbf{D}_{ij}=\textbf{X}_i-\textbf{X}_j$ is the derivatives of $\textbf{d}_{ij}$  with respect to  $\textbf{P}$, which hold due to the fact that
\begin{align}
\frac{\partial \textbf{Ax}}{\partial \textbf{x}}=\textbf{A}^H,
\end{align}
\begin{align}
\textbf{A}_{\rm{H}}=\textbf{H}^H\textbf{W}_b\boldsymbol{\Omega}_{\rm{B}}^{-1}\textbf{W}_b^H\textbf{H},
\end{align}
and
\begin{align}
\textbf{A}_{\rm{G}}=\textbf{G}^H\textbf{W}_e\boldsymbol{\Omega}_{\rm{E}}^{-1}\textbf{W}_e^H\textbf{G}.
\end{align}
%The product of the second and third term of the right-hand side in (\ref{X_B}) and (\ref{X_E}) is given by the fact that $\frac{\partial \textbf{x}^H\textbf{Ax}}{\partial \textbf{x}}=(\textbf{A}+\textbf{A}^H)\textbf{x}$. In addition, $\nabla(\cdot)$ denotes the gradient operation.

In order to find a locally optimal  $\textbf{P}$, we first initialize $\textbf{P}$ and $R_s^{a}$,  solve the gradient $\nabla_{\textbf{P}} R_s^{a}(\textbf{P})$, and adjust $\textbf{P}$ according to $\nabla_{\textbf{P}} R_s^{a}(\textbf{P})$. The value of $\textbf{P}$ is updated according to the following iterative formula
\begin{align} \label{updateP}
{{\mathbf{P}}_{k + 1}} = {{\mathbf{P}}_k} + \mu {\nabla _{\mathbf{P}}}R_s^a\left( {\mathbf{P}} \right).
\end{align}
Then, obtain $R_s^{a}$, update $\textbf{P}$ or step size $\mu$ according to the difference between before and after $R_s^{a}$, and repeat the above steps until the termination condition is reached. The steps of this algorithm are summarized as shown in \textbf{Algorithm \ref{alg:2}}.

%\textbf{Proposed Max-ASR-GA Hybrid Precoding Scheme}:\\
%\textbf{Step 1}: Initializes the precoding matrix $\textbf{P}_0=\textbf{V}(:,1)$, and set initial step size $\mu=3$, $k=0$.\\
%\textbf{Step 2}: Substitute the precoding matrix $\textbf{P}_k$ into (\ref{Rsa}) to obtain the ASR $R_s^{a}(\textbf{P}_k)$.\\
%\textbf{Step 3}: Calculate the gradient $\nabla_{\textbf{P}_k} R_s^{a}(\textbf{P}_k)$ according to (\ref{gradient}), and then update the precoding matrix with (\ref{updateP}) to get $\textbf{P}_{k+1}$.\\
%\textbf{Step 4}: Substitute the precoding matrix $\textbf{P}_{k+1}$ into (\ref{Rsa}) to obtain the ASR $R_s^{a}(\textbf{P}_{k+1})$.\\
%\textbf{Step 5}: Judge whether the termination condition $R_s^a\left( {{{\mathbf{P}}_{k + 1}}} \right)\! -\! R_s^a\left( {{{\mathbf{P}}_k}} \right) \!>\! {10^{ - 4}}$ is satisfied. If so, let $k=k+1$, and return to Step 3 for iteration. Otherwise, let $\mu=\mu/3$.\\
%\textbf{Step 6}: Judge whether the termination condition $\mu\!>\!\mu_{min}$ is satisfied. If so, the iteration will terminate and output the hybrid precoding matrix $\textbf{P}=\textbf{P}_{k+1}$. Otherwise, return to Step 3 for iteration.\\
%\textbf{Step 7}: Let $\textbf{f}_i=\frac{1}{\sqrt{N_{\rm{AA}}}}e^{j\textrm{angle}(\textbf{p}_i)}$ as the analog precoder of the $i$-th TAS and $p_i=\|\textbf{p}_i\|$ as the digital precoder of the $i$-th TAS. Now, we complete the design of hybrid precoding.

\begin{algorithm}
	\renewcommand{\algorithmicrequire}{\textbf{Input:}}
	\renewcommand{\algorithmicensure}{\textbf{Output:}}
	\caption{Max-ASR-GA Hybrid Precoding Scheme}
	\label{alg:2}
	\begin{algorithmic}[1]
		\REQUIRE the channel matrix $\textbf{H}$ and $\textbf{G}$, the M-ary constellation, and TASS matrix $\textbf{T}$
		\ENSURE $\textbf{P}$, $\textbf{F}_{\rm{RF}}$, $\textbf{F}_{\rm{BB}}$
		\STATE Initialize step size $\mu=3$, $\mu_{min}=0.01$ and $t=-1$
        \STATE Initialize the precoding matrix $\textbf{P}_0=\textbf{V}(:,1)$
        \STATE Initialize the ASR $R_s^{a}(\textbf{P}_0)$ based on $\textbf{P}_0$ according to (\ref{Rsa})
        \REPEAT
        \REPEAT
        \STATE Let $k=k+1$
        \STATE Update the gradient $\nabla_{\textbf{P}_k} R_s^{a}(\textbf{P}_k)$ according to (\ref{gradient})
        \STATE Update the precoding matrix $\textbf{P}_{k+1}$ according to (\ref{updateP})
        \STATE Update the ASR $R_s^{a}(\textbf{P}_{k+1})$ based on $\textbf{P}_{k+1}$ according to (\ref{Rsa})
        \UNTIL{$R_s^a\left( {{{\mathbf{P}}_{k + 1}}} \right) - R_s^a\left( {{{\mathbf{P}}_k}} \right) \leq {10^{ - 4}}$}
        \STATE Let $\mu=\mu/3$
        \UNTIL{$\mu<\mu_{min}$}
        \STATE Compute $\textbf{f}_i$ and $p_i$ as the $i$-th TAS analog and digital precoders
        \STATE Substitute $\textbf{f}_i$ and $p_i$ into (\ref{F_rf}) and (\ref{F_bb})
        \STATE \textbf{return} $\textbf{P}=\textbf{P}_{k+1}$, $\textbf{F}_{\rm{RF}}$, $\textbf{F}_{\rm{BB}}$
	\end{algorithmic}
\end{algorithm}

\subsection{Extended SDR-AltMin}

The SDR-AltMin algorithm in \cite{Yu2016Alternating} is specially constructed for traditional hybrid MIMO systems with partially connected structure. Below, we extend it to the hybrid SM systems. In terms of \cite{Yu2016Alternating}, the optimization problem of digital and analog hybrid precoder matrices can be casted as
\begin{align} \label{Frf,Fb}   \nonumber
&\mathop{\textrm{minimize}}\limits_{\textbf{F}_{\rm{RF}},\textbf{F}_{\rm{BB}}} \ \|\textbf{F}_{\rm{opt}}-\textbf{F}_{\rm{RF}}\textbf{F}_{\rm{BB}}\|_F^2 \\
&\textrm{subject} \ \textrm{to} \
\left\{
        \begin{array}{lr}
        \textbf{F}_{\rm{RF}}\in \mathcal{A}\\
        \|\textbf{F}_{\rm{RF}}\textbf{F}_{\rm{BB}}\|_F^2=N_{\rm{RF}}
        \end{array}
\right.,
\end{align}
where $\textbf{F}_{\rm{opt}}\in \mathbb{C}^{N_{\rm{AA}}N_{\rm{RF}} \times 1}$ is the optimal FD precoder in terms of singular value decomposition (SVD) criterion, that is, the first right singular vector of $\textbf{H}$, so $\textbf{F}_{\rm{opt}}=[v_1 \ v_2 \ \cdots \ v_{N_{\rm{AA}}N_{\rm{RF}}}]^T$. $\mathcal{A}$ is the feasible set of the analog precoder $\textbf{F}_{\rm{RF}}$, which is a block diagonal matrix satisfying (\ref{F_rf}). Therefore, the power constraint in (\ref{Frf,Fb}) can be rewritten as $\|\textbf{F}_{\rm{RF}}\textbf{F}_{\rm{BB}}\|_F^2=\|\textbf{F}_{\rm{BB}}\|_F^2=N_{\rm{RF}}$. Here, the power constraint of $N_{\rm{RF}}$ can ensure that the expected power constraint of each TAS is 1.

%$\textbf{f}_i$ and $p_i$ can be solved iteratively by using the $\textbf{SDR-AltMin}$ $\textbf{Algorithm}$ in \cite{Yu2016Alternating}.
%
%According to (\ref{Frf,Fb}), Fix analog precoding $\textbf{f}_i$, the solution of digital precoding $p_i$ can be deduced to semidefinite relaxation (SDR) problem as follows
Once $\textbf{F}_{\rm{BB}}$ is known, $\textbf{F}_{\rm{RF}}$ can be obtained by
\begin{align}  \label{Frf} \nonumber
&(\textbf{F}_{\rm{RF}})_{i,l}=\frac{1}{N_{\rm{AA}}} e^{j\arg\{v_ip_l^H\}}, \\
&1\leq i\leq N_{\rm{AA}}N_{\rm{RF}},\  l=\left\lceil\frac{i}{N_{\rm{AA}}}\right\rceil,
\end{align}
Once $\textbf{F}_{\rm{RF}}$ is known, the digital precoder design problem can be converted into a SDP problem as follows
\begin{align}  \label{Fbb-SDP}  \nonumber
&\mathop{\textrm{minimize}}\limits_{\textbf{Y}\in\mathbb{H}^{N_{\rm{RF}}+1}} \ \textrm{Tr}(\textbf{CY}) \\
&\textrm{subject} \ \textrm{to} \
\left\{
             \begin{array}{lr}
             \textrm{Tr}(\textbf{A}_1\textbf{Y})=N_{\rm{RF}} &  \\
             \textrm{Tr}(\textbf{A}_2\textbf{Y})=1\\
             \textbf{Y}\succeq0. &
             \end{array}
\right.
\end{align}
where $\mathbb{H}^{N_{\rm{RF}}+1}$ is the set of $(N_{\rm{RF}}+1)$-dimensional complex hermitian matrices, $\textbf{y}=[\textbf{F}_{\rm{BB}} \ t]^T$ with an auxiliary variable $t$ and $\textbf{Y}=\textbf{yy}^H$,
\begin{align}
\textbf{C}=
\begin{bmatrix}
    \textbf{F}_{\rm{RF}}^H\textbf{F}_{\rm{RF}}  & -\textbf{F}_{\rm{RF}}^H\textbf{F}_{\rm{opt}} \\
    -\textbf{F}_{\rm{opt}}^H\textbf{F}_{\rm{RF}} & \textbf{F}_{\rm{opt}}^H\textbf{F}_{\rm{opt}}
\end{bmatrix},
\end{align}
\begin{align}
\textbf{A}_1=\begin{bmatrix} \textbf{I}_{N_{\rm{RF}}} & \textbf{0} \\ \textbf{0} & 0\end{bmatrix},
\end{align}
and
\begin{align}
\textbf{A}_2=\begin{bmatrix} \textbf{0}_{N_{\rm{RF}}} & \textbf{0} \\ \textbf{0} & 1\end{bmatrix}.
\end{align}
The optimization problem in (\ref{Fbb-SDP}) can be solved by standard convex optimization techniques. Via alternate iteration between (\ref{Frf}) and (\ref{Fbb-SDP}), the corresponding digital precoding $\textbf{F}_{\rm{BB}}$ and analog precoding $\textbf{F}_{\rm{RF}}$ matrices can be obtained.

%After a certain number of iterations between (\ref{Fb}) and (\ref{Frf}), digital and analog precoding $p_i,\textbf{f}_i$ can be obtained.

At Bob, as shown in Fig.~\ref{fig1}, the multiple-antenna receiver of Bob is fully digital structure. To fully exploit the receive spatial diversity gain, the receive beamforming or combiner is required. Similar to the SVD precoder at transmitter,   here the receive combiner is designed as $\textbf{W}_b=[\textbf{w}_1^b,\textbf{w}_2^b,\ldots,\textbf{w}_{N_{RF}}^b]$, where  $\textbf{w}_i^b\!\in\! \mathbb{C}^{N_{b} \times 1},i\!=\!1,2,\ldots,N_{RF}$.
$\textbf{u}_i,i\!=\!1,2,\ldots,N_{RF}$ is the first left singular vector of $\lambda_i$ for $\textbf{h}_i$. To harvest a high channel gain, $\textbf{u}_i$ can be set to be Bob's combiners for the $i$th TAS, so $\textbf{w}_i\!=\!\textbf{u}_i$. Therefore, the receive beamforming vector at Bob is
\begin{align} \label{W}
\textbf{W}_b=\left[\textbf{u}_1,\textbf{u}_2,\cdots,\textbf{u}_{N_{RF}}\right]
\end{align}
Until now, we complete the design of all precoder schemes and receive beamformer. In particular, the receive beamforming vector in (\ref{W}) will be  adopted at  legitimate receiver in this paper regardless of the kind of precoders and TASSs.

%$\textbf{v}_i^T$ is the first row of $\textbf{V}_i$
%
%\begin{align}
%\textbf{f}_i=\frac{1}{\sqrt{N_{AA}}}e^{j\textrm{angle}(\textbf{v}_i)}
%\end{align}
%
%$\textbf{u}_i^T$ is the first row of $\textbf{U}_i$
%
%$\textbf{W}_b=[\textbf{w}_1,\textbf{w}_2,\ldots,\textbf{w}_{N_{RF}}]$
%
%\begin{align}
%\textbf{w}_i=\frac{1}{\sqrt{N_{b}}}e^{j\textrm{angle}(\textbf{u}_i)}
%\end{align}

\section{Proposed Transmit Antenna Subarray Selection Methods }
In traditional SM systems, transmit antenna selection  is an important technique to improve the system performance.  For hybrid SM, TAS evolves towards TASS. In this section, three TASS methods are proposed for secure hybrid SM systems: Max-EV, Max-ASR, and Max-P-SINR-ANSNR TASS methods. At the same time, the leakage-based TASS method for traditional SM systems in \cite{Shu2018two} can also be extended to hybrid SM systems.

\subsection{Proposed Max-P-SINR-ANSNR}
Considering AN is viewed as the useful signal of Eve, the SINR at Eve is defined as ANSNR. If the product of SINR at Bob and ANSNR at Eve is maximized, it is guaranteed that  at least one of SINR at Bob and ANSNR at Eve or both is high.  This will  accordingly improve the SR performance. Since SINR and ANSNR is related to the TASS matrix $\textbf{T}$, the AN is only projected onto the TAS selected by TASS. In order to facilitate the design, we still assume that each TAS only send AN at the beginning, and then the SINR at Bob and ANSNR at Eve  with activated  TAS $i$ at can be expressed as
\begin{align}
&\textrm{SINR}_i=\frac{\beta P\|\textbf{W}_b^H\textbf{HE}_i\textbf{F}_{\rm{RF}}\textbf{F}_{\rm{BB}}\|^2}{(1-\beta) P\|\textbf{W}_b^H\textbf{H}\textbf{F}_{\rm{RF}}\textbf{T}_{\rm{BB}}\textbf{n}\|^2+N_{\rm{RF}}\sigma_b^2},
\end{align}
and
\begin{align}
&\textrm{ANSNR}_i=\frac{\beta P\|\textbf{W}_e^H\textbf{GE}_i\textbf{F}_{\rm{RF}}\textbf{F}_{\rm{BB}}\|^2}{(1-\beta) P\|\textbf{W}_e^H\textbf{G}\textbf{F}_{\rm{RF}}\textbf{T}_{\rm{BB}}\textbf{n}\|^2+N_{\rm{RF}}\sigma_e^2},
\end{align}
respectively, where $i=1,2,\cdots,N_{\rm{RF}}$. Both $\textrm{SINR}_i$ and $\textrm{ANSNR}_i$ need to increase. Thus, if we multiply $\textrm{SINR}_i$ and $\textrm{ANSNR}_i$, their product will form a maximum value at some TASS.
Taking the logarithm of their product yields
\begin{align} \label{P-SINR-ANSNR} \nonumber
&\log_2\left(\textrm{SINR}*\textrm{ANSNR}\right)\\
&=\log_{2}(\textrm{SINR})-\log_2\left(\frac{1}{\textrm{ANSNR}}\right)\\ \nonumber
&\approx R_B-R_E.
\end{align}
Observing (\ref{P-SINR-ANSNR}),  we can find that maximizing their product is approximately equivalent to maximizing the SR. Therefore, the product of  SINR and ANSNR corresponding to each TAS can be obtained independently, and the first $N_t$ largest products are  selected as candidates. It can be expressed as
\begin{align}
f_i=\textrm{SINR}_i\cdot\textrm{ANSNR}_i.
\end{align}
Then, we sorts $f_i$ in a descending order
\begin{align}
\underbrace{f_{\tau_1}\ge f_{\tau_2}\ge \cdots\ge f_{\tau_{N_t}}}_{\text{Set~of~chosen~TASs}}\ge\cdots\ge f_{\tau_{N_{\rm{RF}}}},
\end{align}
where $\left\{\tau_1,\! \tau_2,\cdots,\! \tau_{N_{\rm{RF}}}\right\}$ is a permutation of $\left\{1,\! 2,\cdots,\! N_{RF}\right\}$. The TASs corresponding to the former $N_t$ $f_{\tau_{N_i}}$ are selected as TASs to transmit SM information.

\subsection{Proposed Max-EV}
In general, a high channel gain corresponds to a good quality of channel, and to achieve a high transmission rate. Therefore, selecting $N_t$ transmit antennas with high channel gains chosen from Alice to Bob can substantially enhance the performance of the system. Before selecting the activated TAS by spatial bits, Alice performs SVD on the channel matrix $\textbf{h}_i$ firstly, as $\textbf{h}_i=\textbf{U}_i\boldsymbol{\Sigma}_i\textbf{V}^H_i,i=1,2,\cdots,N_{\rm{RF}}$, and $\lambda_i$ is the maximum singular value of $\boldsymbol{\Sigma}_i$.
$\textbf{v}_i\in \mathbb{C}^{N_{\rm{AA}} \times 1}$ is the first column of $\textbf{V}_i$, i.e., the associated right singular vector of $\lambda_i$ for $\textbf{h}_i$. And $\textbf{u}_i\in \mathbb{C}^{N_{\rm{AA}} \times 1}$ is the first column of $\textbf{U}_i$, i.e.,  the corresponding left singular vector of $\lambda_i$ for $\textbf{h}_i$.

The maximum singular value $\lambda_i$ means  the largest channel gain, i.e., the best quality of channel. Therefore, Alice sorts $\lambda_i$ in a descending order and renames them as: $\lambda_{\pi_1}$, $\lambda_{\pi_2}$,$\cdots$, $\lambda_{\pi_{N_t}}$,$\cdots$, $\lambda_{\pi_{N_{\rm{RF}}}}$.  Therefore, $\lambda_i$ can be represented as follows
\begin{align} \label{pi}
\underbrace{\lambda_{\pi_1}\ \geq\ \lambda_{\pi_2}\ \geq\ \cdots\ \geq\ \lambda_{\pi_{N_t}}}\limits_{N_t \ \rm{selected \ TASs}}\ \geq\cdots\geq\ \lambda_{\pi_{N_{\rm{RF}}}},
\end{align}
where
\begin{align}
\lambda_{\pi_1}=\max\{\lambda_1,\lambda_2,\cdots,\lambda_{N_{\rm{RF}}}\},
\end{align}
and
\begin{align}
\lambda_{\pi_{N_{\rm{RF}}}}=\min\{\lambda_1,\lambda_2,\cdots,\lambda_{N_{\rm{RF}}}\}.
\end{align}
The former $N_t$ subchannels in (\ref{pi}) are selected to transmit information. We map the spatial position of $\lambda_{\pi_1},\lambda_{\pi_2},\cdots,\lambda_{\pi_{N_t}}$ into the spatial bits. For example, if $N_{\rm{RF}}=7$ and $N_t=4$, we select $\lambda_{\pi_1},\lambda_{\pi_2},\lambda_{\pi_3}$ and $\lambda_{\pi_4}$ to transmit information, the spatial bits $00$ denote that the $\pi_1$th TAS is activated to send  modulation symbol.

Using the above SVD-based TASS method, in the TASS matrix $\textbf{T}=\textrm{diag}[\textbf{t}_1,\textbf{t}_2,\cdots,\textbf{t}_{N_{\rm{RF}}}]$, only $\textbf{t}_{\pi_1}$, $\textbf{t}_{\pi_2},\cdots,\textbf{t}_{\pi_{N_t}}$ are equal to the identity matrix $\textbf{I}_{N_{\rm{AA}}}$, and the remainders are  all-zero matrices $\textbf{0}$.

\subsection{Proposed Max-ASR}
The two previous proposed TASS methods, Max-EV and Max-P-SINR-ANSNR, are extremely low-complexity. However, this low-complexity is achievable at the expense of some SR performance loss, which will be verified in the next simulation section.  Thus, there is still a substantial enhancement room  for secrecy performance. In order to further improve the SR performance, in what follows, by selecting TASs,  the problem of directly maximizing SR (Max-SR) be expressed as follows
\begin{align}  \label{Max-SR-TASS}\nonumber
&\max \ R_s(\textbf{T}) \\
&\textrm{subject} \ \textrm{to} \ \textbf{T}\in\left\{\textbf{T}_1,\textbf{T}_2,\ldots,\textbf{T}_Q\right\}.
\end{align}

Considering the complicated expression of SR yield a dramatically high computational complexity, the  ASR  $R_s^{a}=I_0^{B}-I_0^{E}$  in (\ref{Rsa}) will be used to replace the SR in order to reduce the computational complexity. The Max-SR in (\ref{Max-SR-TASS}) is simplified as
\begin{align}  \label{Max-ASR-TASS}\nonumber
&\max \ R_s^{a}(\textbf{T})= \textrm{log}_2\kappa_E(\textbf{T})- \textrm{log}_2\kappa_B(\textbf{T}) \\
&\textrm{subject} \ \textrm{to} \ \textbf{T}\in\left\{\textbf{T}_1,\textbf{T}_2,\ldots,\textbf{T}_Q\right\},
\end{align}
where $\kappa_B$ and $\kappa_E$ are shown in (\ref{kB}) and (\ref{kE}). $\boldsymbol{\Omega}_{\rm{B}}^{-1}(\textbf{T})$ and $\boldsymbol{\Omega}_{\rm{E}}^{-1}(\textbf{T})$ are related to TASS matrix. And since $\textbf{x}=\textbf{E}_i\textbf{F}_{\rm{RF}}\textbf{F}_{\rm{BB}}b_j$, $\textbf{E}_i$ is used to activate one of the selected transmit TASs, $\textbf{d}_{ij}=\textbf{x}_i-\textbf{x}_j$ and $\textbf{d}_{mk}=\textbf{x}_m-\textbf{x}_k$ are also related to TASS matrix. Therefore, (\ref{kB}) and (\ref{kE}) can be rewritten as
\begin{align}  \nonumber
&{\kappa _B}\left( {\mathbf{T}} \right) = \sum\limits_{i = 1}^{{N_t}M} {\sum\limits_{j = 1}^{{N_t}M} } \\
&{\exp }\left[ {\frac{{ - \beta P{\mathbf{d}}_{ij}^H\left( {\mathbf{T}} \right){{\mathbf{H}}^H}{{\mathbf{W}}_b}{\mathbf{\Omega }}_{\text{B}}^{ - 1}\left( {\mathbf{T}} \right){\mathbf{W}}_b^H{\mathbf{H}}{{\mathbf{d}}_{ij}}\left( {\mathbf{T}} \right)}}{4}} \right],
\end{align}
\begin{align}  \nonumber
&{\kappa _E}\left( {\mathbf{T}} \right) = \sum\limits_{i = 1}^{{N_t}M} {\sum\limits_{j = 1}^{{N_t}M} }\\
&{\exp } \left[ {\frac{{ - \beta P{\mathbf{d}}_{ij}^H\left( {\mathbf{T}} \right){{\mathbf{G}}^H}{{\mathbf{W}}_e}{\mathbf{\Omega }}_{\text{E}}^{ - 1}\left( {\mathbf{T}} \right){\mathbf{W}}_e^H{\mathbf{G}}{{\mathbf{d}}_{ij}}\left( {\mathbf{T}} \right)}}{4}} \right],
\end{align}

Since the TASS in (\ref{Max-ASR-TASS}) is related to both channel and transmitted modulation symbols, and it is selected for secrecy performance, its performance is better than that of TASS method considering only channel. The Max-SR method in \cite{Shu2018two} is also select the transmit antenna for ASR. However, the ASR in \cite{Shu2018two} is not affected by the modulation symbol, so when this method is extended to the  hybrid SM system, its performance is inferior to the method in this paper.

%\begin{align}
%&\kappa_B(\textbf{T})=\sum \limits_{i=1}^{N_tM}\sum \limits_{j=1}^{N_tM}\exp\left( \frac{-\beta P\textbf{d}_{ij}^H\textbf{H}^H\textbf{W}_b\boldsymbol{\omega}_{\rm{b}}(\textbf{T})\textbf{W}_b^H\textbf{H}\textbf{d}_{ij}}{4} \right),  \\
%&\kappa_E(\textbf{T})= \sum \limits_{m=1}^{N_tM}\sum \limits_{k=1}^{N_tM}\exp\left( \frac{-\beta P\textbf{d}_{mk}^H\textbf{G}^H\textbf{W}_e\boldsymbol{\omega}_{\rm{e}}(\textbf{T})\textbf{W}_e^H\textbf{G}\textbf{d}_{mk}}{4} \right),
%\end{align}
%where $\boldsymbol{\omega}_{\rm{b}}(\textbf{T})=\boldsymbol{\Omega}_{\rm{B}}^{-1}(\textbf{T})$ $\boldsymbol{\omega}_{\rm{e}}(\textbf{T})=\boldsymbol{\Omega}_{\rm{E}}^{-1}(\textbf{T})$
\subsection{Extended leakage}
The leakage method for traditional SM systems in \cite{Shu2018two} can also be extended to hybrid spatial modulation systems.
The signal-to-leakage-and-noise ratio (SLNR) for the $n$-th channel is defined as
\begin{align}
&\textrm{SLNR}_n\!=\!\frac{\beta P{\rm tr}\left[(\textbf{W}_b^H\textbf{h}_n\textbf{f}_np_n)(\textbf{W}_b^H\textbf{h}_n\textbf{f}_np_n)^H\right]}{\beta P{\rm \bar{} tr}\left[(\textbf{W}_e^H\textbf{g}_n\textbf{f}_np_n)(\textbf{W}_e^H\textbf{g}_n\textbf{f}_np_n)^H\right]\!+\!N_{\rm {RF}}\sigma_b^2},\!
\end{align}
Then,  the SLNR values per TAS are firstly computed  and   simply arranged in a descending order. The former $N_t$ TASs can be selected as TASs. Although this method has low-complexity, it  considers the leakage from the channel from Alice to Bob into the channel from Alice to Eve.  Actually, a small leakage implies a low information leakage. In other words, a good SR performance can be available.

\subsection{Computational Complexity Analysis}
In this subsection, the approximate computational complexities of the three proposed TASS methods and the extended leakage method in \cite{Shu2018two} are analyzed.
According to \cite{boyd2004convex,Golub1996Matrix}, what we know are: \\
(1)~A complex multiplication or division requires 6 floating-point operations (FLOPs), and a complex addition  or  subtraction requires two FLOPs;\\
(2)~For a complex matrix $\textbf{A}\in \mathbb{C}^{m \times n}$, SVD costs $24mn^2+48m^2n+54m^3$ FLOPs;\\
(3)~Multiplying the complex matrices $\textbf{A}$ and $\textbf{B}\in \mathbb{C}^{n \times p}$, costs $8mnp-2mp$ FLOPs.

The unit of computational complexity is FLOPs, which is ignored for convenience in what follows.

For the SINR and ANSNR of $N_{\rm{RF}}$ TASs of the proposed Max-P-SINR-ANSNR, the computational complexities of those fixed terms in the denominators are omitted. The terms $\beta P\|\textbf{W}_b^H\textbf{HE}_i\textbf{F}_{\rm{RF}}\textbf{F}_{\rm{BB}}\|^2$ and $\beta P\|\textbf{W}_e^H\textbf{GE}_i\textbf{F}_{\rm{RF}}\textbf{F}_{\rm{BB}}\|^2$  in the numerators are calculated for each TAS. Therefore, the computational complexity of Max-P-SINR-ANSNR can be approximated as follows
\begin{align}  \nonumber \label{C-SINR}
&\mathcal{C}_{\rm{Max-P-SINR-ANSNR}}=2N_{\rm{RF}}\cdot\\ \nonumber
&~~~~~~~~~~~~\left[8N_{\rm{RF}}N^2-2N_{\rm{RF}}N+8N_{\rm{RF}}^2N-2N_{\rm{RF}}^2\right]\\
&~~~~~~~~~~~~=\mathcal{O}(N_{\rm{RF}}^2N^2),
\end{align}

For the proposed Max-EV, according to the computational complexity of SVD \cite{Golub1996Matrix}, its complexity has the following approximate form
\begin{align}\nonumber \label{C-EV}
\mathcal{C}_{\rm{Max-EV}}&=N_{\rm{RF}}\left[2N_bN_{\rm{AA}}^2+48N_b^2N_{\rm{AA}}+54N_b^3\right]\\ \nonumber
&=\mathcal{O}(N_{\rm{RF}}N_bN_{\rm{AA}}^2)\\
&=\mathcal{O}(NN_bN_{\rm{AA}}),
\end{align}

For the proposed Max-ASR, ignoring the computational complexity of those fixed terms, the computational complexity is partitioned into three parts: (a)  all possible combinations of $\textbf{x}$, (b) the computation of $\boldsymbol{\Omega}_{\rm{B}}$ and $\boldsymbol{\Omega}_{\rm{B}}$, and (c) the computation of ${\kappa _B}\left( {\mathbf{T}} \right)$ and ${\kappa _E}\left( {\mathbf{T}} \right)$. We can easily get the computational complexity of part (a)
\begin{align}
\mathcal{C}_{\textbf{x}}=MN_t(8N^2-2N+1).
\end{align}
The computational complexity of part (b) can be approximated as
\begin{align}\nonumber
\mathcal{C}_{\boldsymbol{\Omega}_{\rm{B}}}&=\mathcal{C}_{\boldsymbol{\Omega}_{\rm{E}}}=4\cdot\\
&(4N_{\rm{RF}}N^2+N_{\rm{RF}}^2N-N_{\rm{RF}}N+N_{\rm{RF}}^2+2N_{\rm{RF}}).
\end{align}
For part (c), we have
\begin{align}
\mathcal{C}_{{\kappa _B}\left( {\mathbf{T}} \right)}=\mathcal{C}_{{\kappa _E}\left( {\mathbf{T}} \right)}=2(MN_t)^2(8N_{\rm{RF}}^2+2N_{\rm{RF}}+1).
\end{align}
To make an exhaustive search of $Q$ possible TASSs, the above three parts need to be repeated $Q=\begin{pmatrix} N_{\rm{RF}} \\ N_t \end{pmatrix}=\frac{N_{\rm{RF}}!}{N_t!(N_{\rm{RF}}-N_t)!}$ times. Therefore,
\begin{align} \label{C-ASR}
\mathcal{C}_{\rm{Max-ASR}}\!=\!\frac{N_{\rm{RF}}!}{N_t!(N_{\rm{RF}}\!-\!N_t)!}\!\left[\mathcal{C}_{\textbf{x}}\!+\!2\left(\mathcal{C}_{\boldsymbol{\Omega}_{\rm{B}}}\!+\!\mathcal{C}_{{\kappa _B}\left( {\mathbf{T}} \right)}\right)\right].
\end{align}

Since their complexities depend  mainly on $N$ and $N_{\rm{RF}}$ with $N=N_{\rm{AA}}N_{\rm{RF}}$, according to $\mathcal{O}(N_{\rm{RF}}^2N^2)$ and $\mathcal{O}(NN_bN_{\rm{AA}})$ in (\ref{C-SINR}) and (\ref{C-EV}), we find that the computational complexity of Max-EV is far less than that of Max-P-SINR-ANSNR. For the Max-ASR method, the complexity order of each search is $\mathcal{O}(NN_{\rm{RF}}^2)$ according to $\mathcal{C}_{\boldsymbol{\Omega}_{\rm{B}}}$. When $N_{\rm{RF}}=N_t+1$, we get $Q=N_{\rm{RF}}$, therefore the Max-ASR and Max-P-SINR-ANSNR have the same magnitude of complexity. However, the $\mathcal{C}_{\textbf{x}}$ and $\mathcal{C}_{{\kappa _B}\left( {\mathbf{T}} \right)}$ of Max-ASR are also large. Therefore, the complexity of Max-ASR is higher than that of Max-P-SINR-ANSNR. In the same manner, in the case of $N_{\rm{RF}}>N_t+1$, we can obtain the same trend, that is, the computational complexities of the three proposed TASS methods are listed in ascending order: Max-EV, Max-P-SINR-ANSNR, and Max-ASR. As  the value of $N_{\rm{RF}}-(N_t+1)$ increases, the complexity of Max-ASR grows far faster than those of Max-EV and Max-P-SINR-ANSNR.

\section{Simulation Results and Analysis}
In this section, we evaluate the performance of these TASS methods and two hybrid precoding methods proposed by us, with the SDR-AltMin method in \cite{Yu2016Alternating} as a performance reference. The system parameters are set as follows: $N_{\rm{AA}}=4$, $\beta=0.01$, and quadrature phase shift keying (QPSK) modulation. For the convenience of simulation, it is assumed that the total transmit power $P=N_t$ W. For the sake of fairness of Bob and Eve, it is assumed that all noise variances in  channels are identical, i.e., $\sigma_b^2=\sigma_e^2$ and  $N_b=N_e=2$.

\begin{figure}[h]
\centerline{\includegraphics[width=0.48\textwidth]{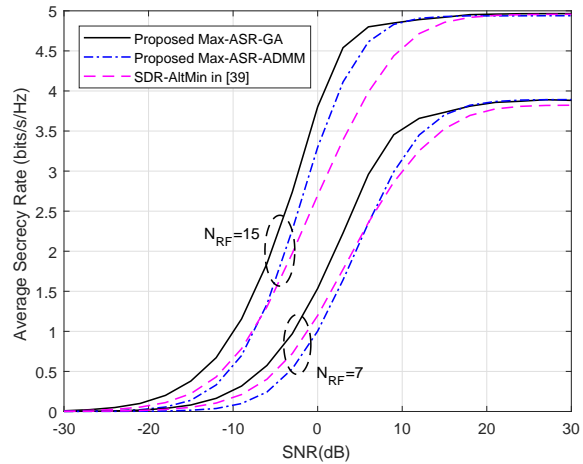}}
\caption{Curves of average SR versus SNR for different hybrid precoding algorithms with $N_{\rm{RF}}=7$ and $N_{\rm{RF}}=15$.}
\label{fig2}
\end{figure}

 When the  Max-EV TASS strategy is adopted in hybrid SM, Fig.~\ref{fig2} demonstrates the average SR performance of the proposed Max-ASR-GA, and proposed Max-ASR-ADMM for $N_{\rm{RF}}=7$ and $N_{\rm{RF}}=15$  with the SDR-AltMin algorithm extended from \cite{Yu2016Alternating} as a performance benchmark. Therefore, $N_t=4$ and $N_t=8$. Considering QPSK modulation is used, when $N_t=4$, the upper bound of the achievable secrecy capacity is $\log_2(MN_t)\!=\!$ 4bits/s/Hz, and when $N_t=4$, the upper bound of the achievable secrecy capacity is $\log_2(MN_t)\!=\!$ 5bits/s/Hz. From  Fig.~\ref{fig2}, it is seen that the SR of the proposed Max-ASR-GA algorithm is much larger than that of extended SDR-AltMin in the low and medium SNR regions.  As the SNR increases, the Max-ASR-GA algorithm can reach the upper bound of the achievable SR with a more rapid rate than  SDR-AltMin regardless of $N_{\rm{RF}}\!=\!7$ or $N_{\rm{RF}}\!=\!15$. This tendency implies that the SR performance of the proposed Max-ASR-GA precoding algorithm is better than that of  SDR-AltMin. The SR  performance of the proposed Max-ASR-ADMM method is in between  the proposed Max-ASR-GA and SDR-AltMin in the medium and high SNR regions, and only slightly lower than SDR-AltMin algorithm in the low SNR region.
 %This is because the Max-ASR-GA and Max-ASR-ADMM algorithms obtain the precoder according to the ASR, while the SDR-AltMin algorithm does not consider the secrecy performance.

\begin{figure}[h]
\centerline{\includegraphics[width=0.48\textwidth]{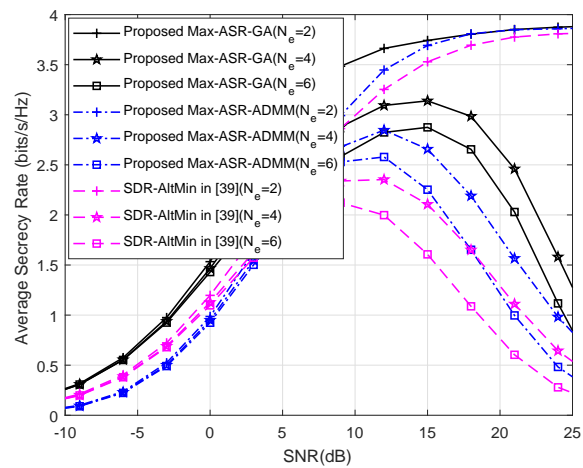}}
\caption{Curves of average SR versus SNR of different hybrid precoding algorithms for  different values of $N_e$ with $N_{\rm{RF}}=7$ and $N_b=2$.}
\label{fig3}
\end{figure}

\begin{figure}[h]
\centerline{\includegraphics[width=0.48\textwidth]{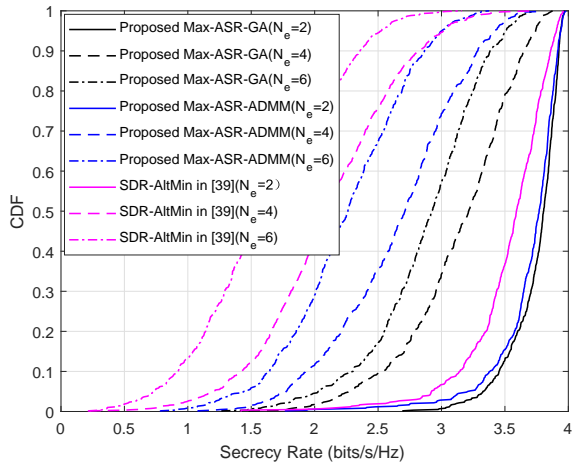}}
\caption{Curves of CDF of SR achieved by different hybrid precoding algorithms for different values of $N_e$ with SNR=15dB, $N_{\rm{RF}}=7$ , and  $N_b=2$.}
\label{fig8}
\end{figure}

Fig.~\ref{fig3} plots the average SR performance achieved by the above the above three hybrid precoding algorithms by fixing $N_{\rm{RF}}\!=\!7$ and $N_b\!=\!2$, and only changing $N_e$.
When $N_b\!=$ $N_e\!=\!2$, as SNR increases, the SR curves increase until it reaches a certain SNR, and the corresponding SR performance is close to the upper bound. However, when $N_e$ is larger than $N_b$, the SR will descend when the SNR is beyond some thresholds.  As $N_e$ increases, the SR curves can achieve the corresponding maximum SRs, and then  decreases. As $N_e$ increases from 4 to 6, the corresponding SNR values of the maximum SRs reached by the three algorithms decreases by about 2.5dB. This is because when SNR is very high, both Bob and Eve have very good quality of channels, while Eve has a larger number of receive antennas than Bob, which is the worst situation, so the SR performance begins to decline. In addition, it can be seen that in the case of $N_e>N_b$, the SR peak of Max-ASR-GA algorithm reaches around 15dB, while the SR peaks of Max-ASR-ADMM and SDR-AltMin algorithms reaches around 12dB and 10dB, respectively. This also shows how well the three algorithms perform.

Fig.~\ref{fig8} shows the CDF curves of the tree precoding algorithms for the different numbers of eavesdropper's antennas when the SNR = 15dB.  In this situation, from  Fig.~\ref{fig8}, we have the same descending trend in SR performance as Fig.~\ref{fig3}: Max-ASR-GA, Max-ASR-ADMM and SDR-AltMin.

\begin{figure}[h]
\centerline{\includegraphics[width=0.48\textwidth]{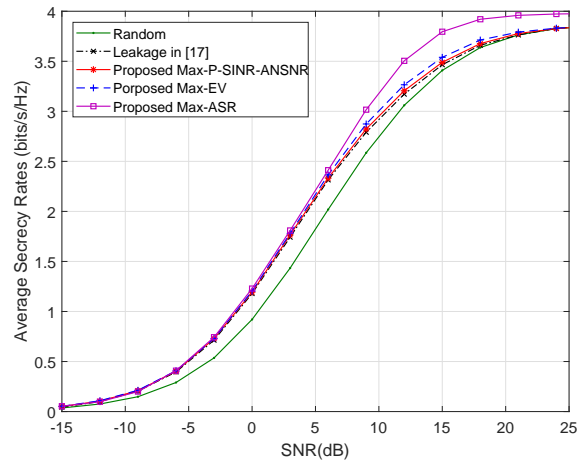}}
\caption{Curves of average SR versus SNR for five TASS methods with $N_{\rm{RF}}=7$ and $N_t=4$.}
\label{fig4}
\end{figure}

Now, we will fix the hybrid precoding method, and make a performance comparison of five TASS strategies. Fig.~\ref{fig4} demonstrates the curves of average SR versus SNR of five TASS methods when $N_{\rm{RF}}=7$ and $N_t=4$. From Fig.~\ref{fig4} it can be clearly seen  that the proposed Max-ASR TASS strategy is  the best one, whereas the  random method is the lowest one, and the remaining methods, such as the proposed Max-EV TASS method and the leakage method in \cite{Shu2018two}, are close to the SR of the Max-ASR method in the low SNR region.  In summary, their SR performance has a descending tendency as follows: Max-ASR, Max-P-SINR-ANSNR, Max-EV, leakage and random.
\begin{figure}[h]
\centerline{\includegraphics[width=0.48\textwidth]{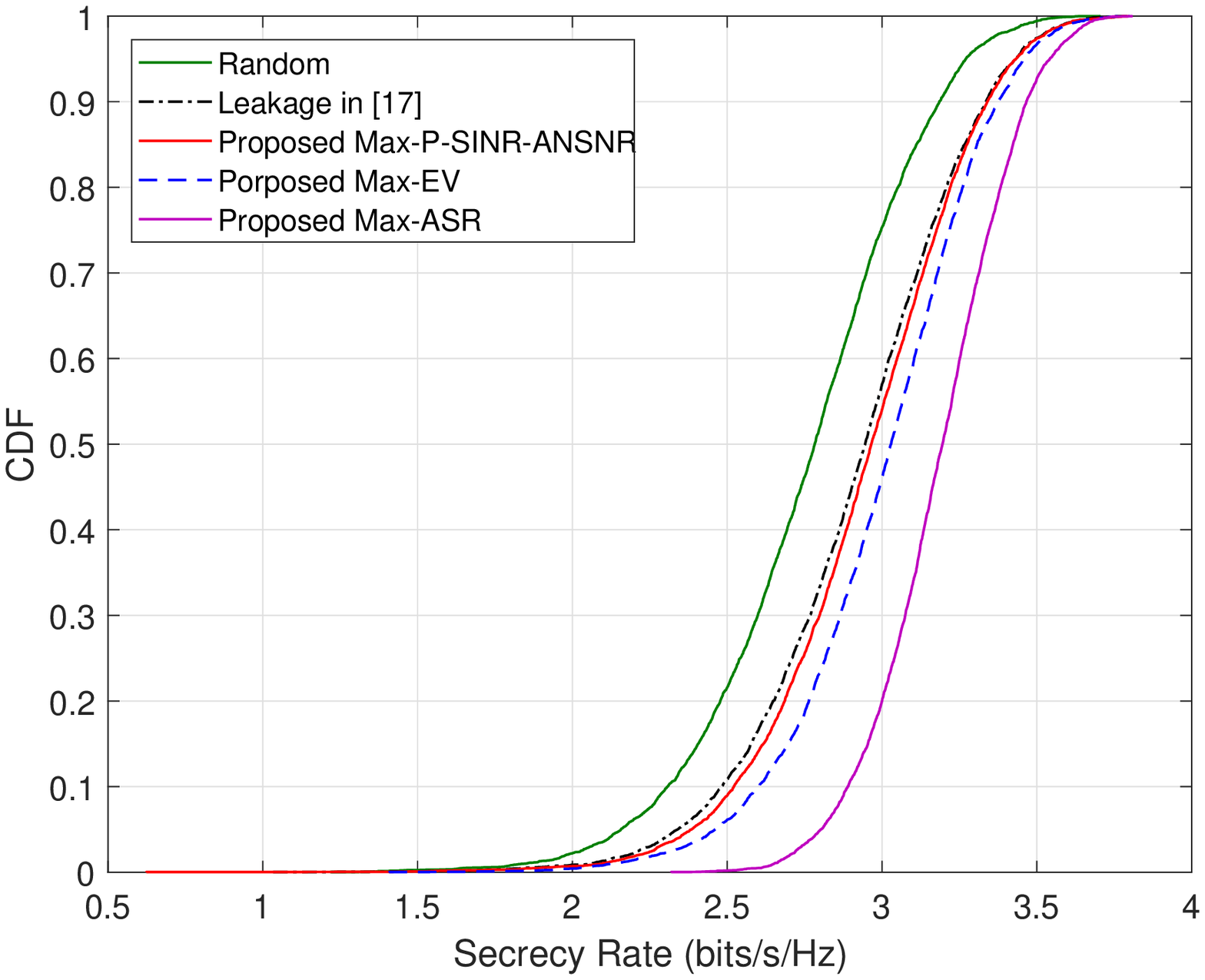}}
\caption{Curves of CDF of SR for SNR=10dB with $N_{\rm{RF}}=7$ and $N_t=4$.}
\label{fig5}
\end{figure}

Fig.~\ref{fig5} illustrates the cumulative distribution function (CDF) curves of SRs of these five TASS methods when $N_{\rm{RF}}=7$  and SNR=10dB. From Fig.~\ref{fig5}, it can be clearly seen that at SNR=10dB, the CDF of  proposed Max-ASR method locates to the most right and that of random is shifted to the most left. In other words, in terms of SR,  the former is the best one, the latter is the worse one,  and the  proposed Max-EV and Max-P-SINR-ANSNR method are in between Max-ASR  and random.

\begin{figure}[h]
\centerline{\includegraphics[width=0.48\textwidth]{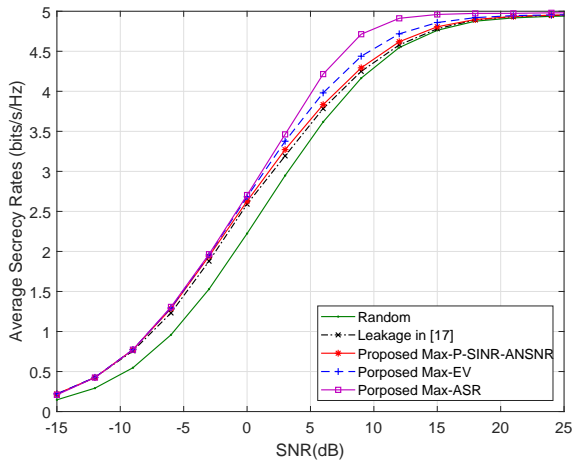}}
\caption{Curves of average SR versus SNR for five TASS methods with $N_{\rm{RF}}=15$ and $N_t=8$.}
\label{fig6}
\end{figure}

\begin{figure}[h]
\centerline{\includegraphics[width=0.48\textwidth]{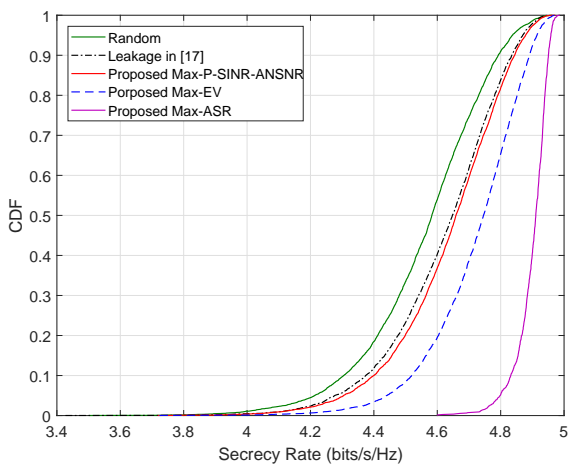}}
\caption{Curves of CDF of SR with SNR=12dB, $N_{\rm{RF}}=15$, and $N_t=8$.}
\label{fig7}
\end{figure}

When $N_{RF}=15$ and $N_t=\lfloor N_{\rm{RF}}\rfloor_2=8$, Fig.~\ref{fig6} plots the average curves of SR versus SNR for five TASS methods. Obviously, the five TASS methods with $N_t=8$ in Fig.~\ref{fig6} has the same trend as that  in Fig.~\ref{fig4} with $N_t=4$ . Fig.~\ref{fig7}shows the CDF curves of SR with $N_t=8$ for SNR=12dB. The five TASS methods has a similar SR performance tendency as Fig.~\ref{fig7}.

\section{Conclusion}
In this paper, we have made a comprehensive investigation of precoding and TASS methods concerning  hybrid  SM. In such a architecture,  the first part of bitstream is transmitted by APM symbol, and the second part of bitstream is carried by selecting a TAS in the partially-connected structure rather than a single transmit antenna. Considering the physical-layer security, three TASS methods, Max-P-SINR-ANSNR, Max-ASR, and Max-EV, were proposed, and at the same time the leakage-based TASS is extended from FD SM to hybrid SM. Particularly, two hybrid precoding algorithms were also proposed:  Max-ASR-GA and Max-ASR-ADM. Simulation results show that the proposed  TASS strategies has an  ascending order in SR: random, leakage, Max-P-SINR-ANSNR, Max-EV, and Max-ASR. Compared with Max-ASR, with dramatically low-complexity,  Max-P-SINR-ANSNR and  Max-EV make a good balance between SR performance and computational complexity. For the hybrid precoding, the proposed Max-ASR-GA and Max-ASR-ADMM hybrid precoding algorithm performs better than existing SDR-AltMin in terms of SR performance.

\vspace{12pt}

\bibliographystyle{IEEEtran}
\bibliography{IEEEabrv,ref}

% Generated by IEEEtran.bst, version: 1.13 (2008/09/30)
\begin{thebibliography}{10}
\providecommand{\url}[1]{#1}
\csname url@samestyle\endcsname
\providecommand{\newblock}{\relax}
\providecommand{\bibinfo}[2]{#2}
\providecommand{\BIBentrySTDinterwordspacing}{\spaceskip=0pt\relax}
\providecommand{\BIBentryALTinterwordstretchfactor}{4}
\providecommand{\BIBentryALTinterwordspacing}{\spaceskip=\fontdimen2\font plus
\BIBentryALTinterwordstretchfactor\fontdimen3\font minus
  \fontdimen4\font\relax}
\providecommand{\BIBforeignlanguage}[2]{{%
\expandafter\ifx\csname l@#1\endcsname\relax
\typeout{** WARNING: IEEEtran.bst: No hyphenation pattern has been}%
\typeout{** loaded for the language `#1'. Using the pattern for}%
\typeout{** the default language instead.}%
\else
\language=\csname l@#1\endcsname
\fi
#2}}
\providecommand{\BIBdecl}{\relax}
\BIBdecl

\bibitem{Foschini2010Layered}
G.~J. Foschini, ``Layered space-time architecture for wireless communication in
  a fading environment when using multi-element antennas,'' \emph{Bell Labs
  Tech. J.}, vol.~1, no.~2, pp. 41--59, Aug. 1996.

\bibitem{Yu2015Power}
X.~Yu, S.~H. Leung, B.~Wu, and Y.~Rui, ``Power control for space-time-coded
  {MIMO} systems with imperfect feedback over joint transmit-receive-correlated
  channel,'' \emph{IEEE Trans. Veh. Technol.}, vol.~64, no.~6, pp. 2489--2501,
  Jun. 2015.

\bibitem{Renzo2011Spatial}
M.~D. Renzo, H.~Haas, and P.~M. Grant, ``Spatial modulation for
  multiple-antenna wireless systems: A survey,'' \emph{{IEEE} Commun. Mag.},
  vol.~49, no.~12, pp. 182--191, Dec. 2011.

\bibitem{Mesleh2008Spatial}
R.~Y. Mesleh, H.~Haas, S.~Sinanovic, W.~A. Chang, and S.~Yun, ``Spatial
  modulation,'' \emph{{IEEE} Trans. Veh. Technol.}, vol.~57, no.~4, pp.
  2228--2241, Jul. 2008.

\bibitem{Jeganathan2012Spatial}
J.~Jeganathan, A.~Ghrayeb, and L.~Szczecinski, ``Spatial modulation: Optimal
  detection and performance analysis,'' \emph{{IEEE} Commun. Lett.}, vol.~12,
  no.~8, pp. 545--547, Aug. 2012.

\bibitem{Rajashekar2013Antenna}
R.~Rajashekar, K.~V.~S. Hari, and L.~Hanzo, ``Antenna selection in spatial
  modulation systems,'' \emph{IEEE Commun. Lett.}, vol.~17, no.~3, pp.
  521--524, Mar. 2013.

\bibitem{Xia2018AS}
G.~{Xia}, F.~{Shu}, Y.~{Zhang}, J.~{Wang}, S.~{ten Brink}, and J.~{Speidel},
  ``Antenna selection method of maximizing secrecy rate for green secure
  spatial modulation,'' \emph{IEEE Trans. Green Commun. Netw.}, vol.~3, no.~2,
  pp. 288--301, Jun. 2019.

\bibitem{Xia2019AN}
G.~{Xia}, Y.~{Lin}, T.~{Liu}, F.~{Shu}, and L.~{Hanzo}, ``Transmit antenna
  selection and beamformer design for secure spatial modulation with rough
  {CSI} of {E}ve,'' 2019. [Online]. Available:
  https://arxiv.org/abs/1905.10088.

\bibitem{Jin2015Linear}
S.~R. Jin, W.~C. Choi, J.~H. Park, and D.~J. Park, ``Linear precoding design
  for mutual information maximization in generalized spatial modulation with
  finite alphabet inputs,'' \emph{IEEE Commun. Lett.}, vol.~19, no.~8, pp.
  1323--1326, Aug. 2015.

\bibitem{Yang2011Transmitter}
L.~L. Yang, ``Transmitter preprocessing aided spatial modulation for
  multiple-input multiple-output systems,'' in \emph{Proc. IEEE Veh. Technol.
  Conf.}, Yokohama, Japan, May 2011, pp. 1--5.

\bibitem{shu2019high}
F.~Shu, X.~Liu, G.~Xia, T.~Xu, J.~Li, and J.~Wang, ``High-performance power
  allocation strategies for secure spatial modulation,'' \emph{IEEE Trans. Veh.
  Technol.}, vol.~68, no.~5, pp. 5164--5168, May. 2019.

\bibitem{xia2018power}
G.~{Xia}, L.~{Jia}, Y.~{Qian}, F.~{Shu}, Z.~{Zhuang}, and J.~{Wang}, ``Power
  allocation strategies for secure spatial modulation,'' \emph{IEEE Syst. J.},
  vol.~13, no.~4, pp. 3869--3872, Dec. 2019.

\bibitem{Aghdam2015On}
S.~R. Aghdam, T.~M. Duman, and M.~D. Renzo, ``On secrecy rate analysis of
  spatial modulation and space shift keying,'' in \emph{Proc. IEEE Int. Black
  Sea Conf. Commun. Netw.}, Constanta, Romania, May. 2015, pp. 63--67.

\bibitem{Wang2015Secrecy}
L.~Wang, S.~Bashar, Y.~Wei, and R.~Li, ``Secrecy enhancement analysis against
  unknown eavesdropping in spatial modulation,'' \emph{IEEE Commun. Lett.},
  vol.~19, no.~8, pp. 1351--1354, Aug. 2015.

\bibitem{Wang2016Spatial}
X.~Wang, X.~Wang, and L.~Sun, ``Spatial modulation aided physical layer
  security enhancement for fading wiretap channels,'' in \emph{Proc. IEEE Int.
  Conf. Wirel. Commun. Signal Process.}, Yangzhou, China, Oct. 2016, pp. 1--5.

\bibitem{Liu2017Secure}
C.~Liu, L.~L. Yang, and W.~Wang, ``Secure spatial modulation with a full-duplex
  receiver,'' \emph{IEEE Wireless Commun. Lett.}, vol.~6, no.~6, pp. 838--841,
  Dec. 2017.

\bibitem{Shu2018two}
F.~Shu, Z.~Wang, R.~Chen, Y.~Wu, and J.~Wang, ``Two high-performance schemes of
  transmit antenna selection for secure spatial modulation,'' \emph{IEEE Trans.
  Veh. Technol.}, vol.~67, no.~9, pp. 8969--8973, Sept. 2018.

\bibitem{zhao2017artificial}
N.~Zhao, Y.~Cao, F.~R. Yu, Y.~Chen, M.~Jin, and V.~C. Leung, ``Artificial noise
  assisted secure interference networks with wireless power transfer,''
  \emph{{IEEE} Trans. Veh. Technol.}, vol.~67, no.~2, pp. 1087--1098, Feb.
  2017.

\bibitem{zheng2015multi}
T.-X. Zheng, H.-M. Wang, J.~Yuan, D.~Towsley, and M.~H. Lee, ``Multi-antenna
  transmission with artificial noise against randomly distributed
  eavesdroppers,'' \emph{{IEEE} Trans. Commun.}, vol.~63, no.~11, pp.
  4347--4362, Nov. 2015.

\bibitem{chen2016survey}
X.~Chen, D.~W.~K. Ng, W.~H. Gerstacker, and H.-H. Chen, ``A survey on
  multiple-antenna techniques for physical layer security,'' \emph{IEEE Commun.
  Surv. Tutor.}, vol.~19, no.~2, pp. 1027--1053, Jun. 2017.

\bibitem{hu2016robust}
J.~Hu, F.~Shu, and J.~Li, ``Robust synthesis method for secure directional
  modulation with imperfect direction angle,'' \emph{IEEE Commun. Lett.},
  vol.~20, no.~6, pp. 1084--1087, Jun. 2016.

\bibitem{shu2016robust}
F.~Shu, X.~Wu, J.~Li, R.~Chen, and B.~Vucetic, ``Robust synthesis scheme for
  secure multi-beam directional modulation in broadcasting systems,''
  \emph{IEEE Access}, vol.~4, pp. 6614--6623, Oct. 2016.

\bibitem{Zhou2019UAV}
X.~{Zhou}, Q.~{Wu}, S.~{Yan}, F.~{Shu}, and J.~{Li}, ``{UAV}-enabled secure
  communications: Joint trajectory and transmit power optimization,''
  \emph{IEEE Trans. Veh. Technol.}, vol.~68, no.~4, pp. 4069--4073, Apr. 2019.

\bibitem{zou2016physical}
Y.~Zou, ``Physical-layer security for spectrum sharing systems,'' \emph{{IEEE}
  Trans. Wireless Commun.}, vol.~16, no.~2, pp. 1319--1329, Feb. 2017.

\bibitem{Wu2015Secret}
F.~Wu, L.~L. Yang, W.~Wang, and Z.~Kong, ``Secret precoding-aided spatial
  modulation,'' \emph{IEEE Commun. Lett.}, vol.~19, no.~9, pp. 1544--1547,
  Sept. 2015.

\bibitem{Wu2015Secure}
F.~Wu, D.~Chen, L.~L. Yang, and W.~Wang, ``Secure wireless transmission based
  on precoding-aided spatial modulation,'' in \emph{Proc. IEEE Global Commun.
  Conf.}, San Diego, CA, USA, Dec. 2015, pp. 1--6.

\bibitem{Wu2016Transmitter}
F.~Wu, R.~Zhang, L.~L. Yang, and W.~Wang, ``Transmitter precoding aided spatial
  modulation for secrecy communications,'' \emph{IEEE Trans. Veh. Technol.},
  vol.~65, no.~1, pp. 467--471, Jan. 2016.

\bibitem{Chen2016Secure}
Y.~Chen, W.~Li, Z.~Zhao, M.~Meng, and B.~Jiao, ``Secure multiuser {MIMO}
  downlink transmission via precoding-aided spatial modulation,'' \emph{IEEE
  Commun. Lett.}, vol.~20, no.~6, pp. 1116--1119, Jun. 2016.

\bibitem{Gao2016Energy}
X.~Gao, L.~Dai, S.~Han, I.~Chih-Lin, and R.~W.~H. Jr, ``Energy-efficient hybrid
  analog and digital precoding for {mmWave} {MIMO} systems with large antenna
  arrays,'' \emph{{IEEE} J. Select. Areas Commun.}, vol.~34, no.~4, pp.
  998--1009, Apr. 2016.

\bibitem{shu2018low}
F.~Shu, Y.~Qin, T.~Liu, L.~Gui, Y.~Zhang, J.~Li, and Z.~Han, ``Low-complexity
  and high-resolution doa estimation for hybrid analog and digital massive
  {MIMO} receive array,'' \emph{{IEEE} Trans. Commun.}, vol.~66, no.~6, pp.
  2487--2501, Jun. 2018.

\bibitem{Cui2016Hybrid}
Y.~Cui, X.~Fang, and Y.~Li, ``Hybrid spatial modulation beamforming for
  {mmWave} railway communication systems,'' \emph{{IEEE} Trans. Veh. Technol.},
  vol.~65, no.~12, pp. 9597--9606, Dec. 2016.

\bibitem{Y2017Hybrid}
M.~Y{\"u}zge{\c{c}}cio{\u{g}}lu and E.~Jorswieck, ``Hybrid beamforming with
  spatial modulation in multi-user massive {MIMO} {mmWave} networks,'' in
  \emph{Proc. IEEE Int. Symp. Person. Indoor Mobile Radio Commun.}\hskip 1em
  plus 0.5em minus 0.4em\relax Montreal, QC, Canada: IEEE, Oct. 2017, pp. 1--6.

\bibitem{Sheng2018Macro}
S.~Luo, X.~T. Tran, W.~Wang, K.~C. Teh, and K.~H. Li, ``Macro spatial
  modulation for uplink {mmWave} communication systems,'' in \emph{Proc. IEEE
  Globecom Workshops}.\hskip 1em plus 0.5em minus 0.4em\relax Singapore: IEEE,
  Dec. 2017, pp. 1--5.

\bibitem{lee2017adaptive}
M.~C. Lee and W.~H. Chung, ``Adaptive multimode hybrid precoding for
  {single-RF} virtual space modulation with analog phase shift network in
  {MIMO} systems,'' \emph{{IEEE} Trans. Wireless Commun.}, vol.~16, no.~4, pp.
  2139--2152, Apr. 2017.

\bibitem{He2017Spectral}
L.~He, J.~Wang, and J.~Song, ``Spectral-efficient analog precoding for
  generalized spatial modulation aided {mmWave} {MIMO},'' \emph{{IEEE} Trans.
  Veh. Technol.}, vol.~66, no.~10, pp. 9598--9602, Oct. 2017.

\bibitem{lu2018low}
Z.~Lu, L.~He, J.~Wang, and J.~Song, ``Low complexity hybrid precoding algorithm
  for gensm aided {mmWave} {MIMO} systems,'' in \emph{Proc. IEEE Int. Wirel.
  Commun. Mob. Comput. Conf.}\hskip 1em plus 0.5em minus 0.4em\relax Limassol,
  Cyprus: IEEE, Jun. 2018, pp. 412--417.

\bibitem{he2018spatial}
L.~He, J.~Wang, and J.~Song, ``Spatial modulation for more spatial
  multiplexing: {RF-chain-limited} generalized spatial modulation aided
  {MM-Wave} {MIMO} with hybrid precoding,'' \emph{{IEEE} Trans. Commun.},
  vol.~66, no.~3, pp. 986--998, Mar. 2018.

\bibitem{Aghdam2017Joint}
S.~R. Aghdam and T.~M. Duman, ``Joint precoder and artificial noise design for
  {MIMO} wiretap channels with finite-alphabet inputs based on the {Cut-Off}
  rate,'' \emph{IEEE Trans. Wireless Commun.}, vol.~16, no.~6, pp. 3913--3923,
  Jun. 2017.

\bibitem{Yu2016Alternating}
X.~Yu, J.~C. Shen, J.~Zhang, and K.~Letaief, ``Alternating minimization
  algorithms for hybrid precoding in millimeter wave {MIMO} systems,''
  \emph{{IEEE} J. Sel. Top. Sign. Proces.}, vol.~10, no.~3, pp. 485--500, Apr.
  2016.

\bibitem{boyd2011distributed}
S.~Boyd, N.~Parikh, E.~Chu, B.~Peleato, J.~Eckstein \emph{et~al.},
  ``Distributed optimization and statistical learning via the alternating
  direction method of multipliers,'' \emph{Found. Trends Mach. Learn.}, vol.~3,
  no.~1, pp. 1--122, 2011.

\bibitem{boyd2004convex}
S.~Boyd and L.~Vandenberghe, \emph{Convex optimization}.\hskip 1em plus 0.5em
  minus 0.4em\relax Cambridge university press, 2004.

\bibitem{Golub1996Matrix}
G.~Golub and C.~F. Van~Loan, \emph{Matrix computations. 3rd ed.}\hskip 1em plus
  0.5em minus 0.4em\relax The Johns Hopkins University Press, 1996.

\end{thebibliography}

\end{document}